\documentclass[%
reprint,
superscriptaddress,
nofootinbib,
amsmath,
amssymb,
aps,
prl,
]{revtex4-2}

\usepackage[normalem]{ulem}
\usepackage{graphicx}
\usepackage{dcolumn}
\usepackage{bm}
\usepackage{orcidlink}
\usepackage{microtype}
\usepackage{hyperref}
\hypersetup{
  colorlinks=true,
  citecolor=blue,
  linkcolor=blue,
  urlcolor=blue,
  filecolor=blue,
}

\makeatletter

\renewcommand{\section}{%
    \@startsection{section}{1}{\parindent}%
    {0.0\baselineskip}%
    {0pt}%
    {\normalfont\normalsize\itshape}%
}

\def\@runin@to@section#1#2#3{#1#2#3---}

\makeatother

\usepackage{enumitem}
\usepackage{braket}
\usepackage{mathtools}

\usepackage{caption}
\usepackage{subcaption}
\usepackage{tikz}
\usetikzlibrary{decorations.markings,arrows.meta}
\usepackage[compat=1.1.0]{tikz-feynman}
\tikzfeynmanset{warn luatex=false}
\usepackage{simpler-wick}
\usetikzlibrary{positioning,calc,decorations.markings,decorations.pathmorphing}

\usepackage{tcolorbox}
\tcbuselibrary{minted,breakable,xparse,skins}

\definecolor{bg}{gray}{0.95}
\DeclareTCBListing{mintedbox}{O{}m!O{}}{%
  breakable=true,
  listing engine=minted,
  listing only,
  minted language=#2,
  minted style=default,
  minted options={%
    linenos,
    gobble=0,
    breaklines=true,
    breakafter=,,
    fontsize=\small,
    numbersep=8pt,
    #1},
  boxsep=0pt,
  left skip=0pt,
  right skip=0pt,
  left=25pt,
  right=0pt,
  top=3pt,
  bottom=3pt,
  arc=5pt,
  leftrule=0pt,
  rightrule=0pt,
  bottomrule=2pt,
  toprule=2pt,
  colback=bg,
  colframe=orange!70,
  enhanced,
  overlay={%
    \begin{tcbclipinterior}
    \fill[orange!20!white] (frame.south west) rectangle ([xshift=20pt]frame.north west);
    \end{tcbclipinterior}},
  #3}

\newcommand{\eps}{\varepsilon}

\renewcommand{\epsilon}{\eps}

\usepackage{bbm}

\DeclarePairedDelimiter\abs{\lvert}{\rvert}%
\DeclarePairedDelimiter\norm{\lVert}{\rVert}%

\makeatletter
\let\oldabs\abs
\def\abs{\@ifstar{\oldabs}{\oldabs*}}
\let\oldnorm\norm
\def\norm{\@ifstar{\oldnorm}{\oldnorm*}}
\makeatother

\newcounter{prlrunin}

\begin{document}


\title{Decoherence and the Reemergence of Coherence From a Superconducting ``Horizon''}

\author{Eric J. Sung\orcidlink{0000-0003-4437-1068}}
\email{jsung1@arizona.edu}
\affiliation{Program in Applied Mathematics, University of Arizona, Tucson, AZ 85721, USA}

\author{Charles A. Stafford\orcidlink{0000-0002-7190-1696}}
\email{stafforc@arizona.edu}
\affiliation{Department of Physics, University of Arizona, Tucson, AZ 85721, USA}
\affiliation{Program in Applied Mathematics, University of Arizona, Tucson, AZ 85721, USA}





\date{\today}

\begin{abstract}
In a recent paper \cite{danielsonBlackHolesDecohere2022}, Danielson et al. demonstrated that the mere presence of a black hole causes universal decoherence of quantum superpositions (dubbed the DSW decoherence). 
We analyze decoherence in a superconducting analogue \cite{manikandanAndreevReflectionsQuantum2017} of the event horizon of a late-stage black hole, where Andreev reflection plays the role of Hawking radiation. We consider a normal metal interferometer threaded by an Aharonov-Bohm flux, where one of the arms of the interferometer is coupled to a superconductor by a tunnel coupling of varying strength. At absolute zero temperature and for weak coupling, we find that the scattering states of the interferometer are decohered by Andreev reflection, a nontrivial manifestation of the proximity effect analogous to DSW decoherence from the event horizon of a black hole. However, for increasing coupling strength to the superconductor, we find a reemergence of coherence via resonant tunneling through Andreev bound states.  This suggests the existence of an analogue gravitational phenomenon wherein transmission mediated by virtual Hawking radiation leads to a reemergence of coherence in an interferometer placed near a black hole's event horizon. 

\end{abstract}

\maketitle



Physical analogies reveal that the same mathematical structures and emergent phenomena can arise in systems with very different microscopic origins. This idea has seen popularity in analogue gravity \cite{barceloAnalogueGravityField2001,visserAnalogueModelsGravity2002,barceloAnalogueGravity2011,faccioAnalogueGravityPhenomenology2013,almeidaAnalogueGravityHawking2023}, where condensed matter and quantum optical platforms reproduce hallmark features of curved spacetime physics, such as effective horizons \cite{unruhExperimentalBlackHoleEvaporation1981,unruhSonicAnalogueBlack1995,volovikSuperfluid3HeBGravity1990,volovikLifshitzTransitionsTypeII2017} and Hawking-like radiation \cite{hawkingParticleCreationBlack1975,hawkingBreakdownPredictabilityGravitational1976}. Following the discovery of the Unruh effect \cite{fullingNonuniquenessCanonicalField1973,daviesScalarProductionSchwarzschild1975,unruhNotesBlackholeEvaporation1976}, Unruh derived its sonic analogue \cite{unruhExperimentalBlackHoleEvaporation1981,unruhSonicAnalogueBlack1995}, one of the earliest analogue gravity systems. Since then, a multitude of different systems have been proposed as analogue gravity testbeds, including fermionic and bosonic superfluids \cite{volovikSuperfluid3HeBGravity1990,volovikSimulationPanleveGullstrandBlack1999,fischerThermalQuasiequilibriumStates2001a,barceloAnalogueGravityBoseEinstein2001,schutzholdQuantumBackreactionDilute2005,manikandanBosonsFallingBlack2018}, Weyl and Dirac materials \cite{volovikBlackHoleHawking2016,volovikLifshitzTransitionsTypeII2017}, trapped ions \cite{tianTestingUpperBound2022}, and interacting bosons on optical lattices \cite{bilokonHilbertSpaceBlack2026}. Some of these analogue platforms have also seen experimental confirmation such as analogue Hawking radiation using Bose-Einstein condensates \cite{steinhauerObservationQuantumHawking2016,munozdenovaObservationThermalHawking2019,kolobovObservationStationarySpontaneous2021,fabbriRampupHawkingRadiation2021}, optical analogues \cite{steinhauerObservationSelfamplifyingHawking2014,droriObservationStimulatedHawking2019}, and superconducting quantum processors \cite{shiQuantumSimulationHawking2023}.

In this Letter, we explore the deep analogy \cite{jacobsonOriginOutgoingBlack1996,manikandanAndreevReflectionsQuantum2017,manikandanBlackHolesAndreev2020} of BCS superconductors \cite{bardeenTheorySuperconductivity1957} and late-stage black holes that have already radiated away at least half their initial entropy  \cite{horowitzBlackHoleFinal2004,haydenBlackHolesMirrors2007,lloydUnitarityBlackHole2014}. More specifically, we identify the solid-state analogue of the newly discovered Danielson, Satishchandran, and Wald (DSW) decoherence for black holes \cite{danielsonBlackHolesDecohere2022,danielsonKillingHorizonsDecohere2023,danielsonLocalDescriptionDecoherence2025} in an Aharonov-Bohm (AB) interferometer fabricated of normal metal coupled on one side to a superconductor. We show that weak coupling to the superconductor suppresses interference due to decoherence induced by Andreev scattering, while intermediate coupling leads to a reemergence of coherence via resonant Andreev processes. The conceptual mapping between superconductor-normal metal (SN) systems and black holes opens the possibility of utilizing superconducting interferometers as platforms for studying horizon-induced decoherence in the laboratory.

\begin{figure}[htbp]
  \centering

  \begin{subfigure}{0.48\columnwidth}
    \centering
    \resizebox{\linewidth}{!}{%
    \begin{tikzpicture}
\path[use as bounding box] (-3.6,-2.6) rectangle (3.0,2.6);
    \node[rectangle,
    draw = white,
    minimum width = 2cm, 
    minimum height = 3cm] (r) at (-3,0) {\textbf{S}};

    \draw[thick] (-2.5,1.8) -- (-2.5,-1.8); 
    \draw[thick] (-3.5,1.8) -- (-2.5,1.8); 
    \draw[thick] (-3.5,-1.8) -- (-2.5,-1.8); 

    \node[rectangle,
    draw = white,
    minimum width = 2cm, 
    minimum height = 3cm] (r) at (3.75,0) {};

    \begin{scope}
      \fill[gray!20,opacity=0.7]
        (-2.5,-1.0)
        arc (270:90:-0.50 and 1.0)
        -- cycle;
    \end{scope}

\begin{scope}[xshift=-1.5cm]
    \coordinate (Ltop) at (-0.75,0.6);
    \coordinate (Lbot) at (-0.75,-0.6);
    \coordinate (Rtop) at (1.0,2.0);
    \coordinate (Rbot) at (1.0,-2.0);
    
    \draw[thick, blue,
        decoration={markings, mark=at position 0.45 with {\arrow{<}}},
        postaction={decorate}]
        (Ltop)
        .. controls (-0.25,1.75) and (1.00,2.00) ..
        (Rtop);
    
    \draw[thick, blue,
        decoration={markings, mark=at position 0.45 with {\arrow{>}}},
        postaction={decorate}]
        (Lbot)
        .. controls (-0.25,-1.75) and (1.00,-2.00) ..
        (Rbot);
    
    \draw[thick, blue,
        decoration={
            markings,
            mark=at position 0.25 with {\arrow[scale=1.2]{>}},
            mark=at position 0.75 with {\arrow[scale=1.2]{>}}
        },
        postaction={decorate}]
        (Rtop)
        .. controls (3.5,1.2) and (3.5,-1.2) ..
        (Rbot);

    \path (Rtop) ++(-0.22,-0.03) coordinate (Tleft);
    \path (Rtop) ++( 0.22,-0.03) coordinate (Tright);
    
    \path (Rtop) ++(-0.85,1.0) coordinate (Ileft);
    \path (Rtop) ++( 0.85,1.0) coordinate (Iright);
    
    \draw[thick]
      (Ileft) .. controls +(0,-0.8) and +(-0.25,0.6) .. (Tleft);
    
    \draw[thick]
      (Tright) .. controls +(0.25,0.6) and +(0,-0.8) .. (Iright);
    
    \node[font=\bfseries] at ($(Rtop)+(0,0.75)$) {I};
    
    \path (Rbot) ++(-0.22,0.03) coordinate (Bleft);
    \path (Rbot) ++( 0.22,0.03) coordinate (Bright);
    
    \path (Rbot) ++(-0.85,-1.0) coordinate (IIleft);
    \path (Rbot) ++( 0.85,-1.0) coordinate (IIright);
    
    \draw[thick]
      (IIleft) .. controls +(0,0.8) and +(-0.25,-0.6) .. (Bleft);
    
    \draw[thick]
      (Bright) .. controls +(0.25,-0.6) and +(0,0.8) .. (IIright);
    
    \node[font=\bfseries] at ($(Rbot)+(0,-0.75)$) {II};
\end{scope}
    
    

    \node[scale=1.5] at (-2.25,0.6) {$\times$};
    \node[scale=1.5] at (-2.25,-0.6) {$\times$};

    \coordinate (first_electron_start) at (-2.35,-0.5);
    \coordinate (new_first_electron_start) at ($(first_electron_start)!0.00!(0.6,-1.85)$);
    \coordinate (end) at (-2.35,0.5);
    
    \draw[thick, red, decoration={markings,
            mark=at position 0.45 with {\arrow[scale=0.8]{>}}},
            postaction={decorate}] (new_first_electron_start) -- (end)
            node[midway, xshift=-4mm, yshift=-1.5mm] {};
            
    \coordinate (first_hole_start) at (-2.15,-0.5);
    \coordinate (new_first_hole_start) at ($(first_hole_start)!0.00!(0.6,-1.85)$);
    \coordinate (end) at (-2.15,0.5);
    
    \draw[thick, blue, decoration={markings,
            mark=at position 0.55 with {\arrow[scale=0.8]{<}}},
            postaction={decorate}] (new_first_hole_start) -- (end)
            node[midway, xshift=-4mm, yshift=-1.5mm] {};

\node (flux) at (-0.5,0) {$\Phi$};

\draw[thick] ($(flux)+(96:2.0)$) arc[start angle=96,end angle=264,radius=2.0];
\draw[thick] ($(flux)+(277:2.0)$) arc[start angle=277,end angle=444,radius=2.0];

\draw[thick] (flux) circle [radius=1.6];
    
    \end{tikzpicture}%
    }

    \caption{}
    \label{fig:general_n_ring_superconductor_diagram} 
  \end{subfigure}
    \begin{subfigure}{0.48\columnwidth}
    \centering
    \resizebox{\linewidth}{!}{%
      \begin{tikzpicture}
------------------------------------------------
\path[use as bounding box] (-3.6,-2.6) rectangle (3.0,2.6);

        \node[rectangle,
        draw = white,
        minimum width = 2cm,
        minimum height = 3cm] (r) at (-3,0) {\textbf{S}};

        \draw[thick] (-2.5,1.8) -- (-2.5,-1.8); 
        \draw[thick] (-3.5,1.8) -- (-2.5,1.8); 
        \draw[thick] (-3.5,-1.8) -- (-2.5,-1.8); 

        \node[rectangle,
        draw = white,
        minimum width = 2cm,
        minimum height = 3cm] (r) at (4.75,0){};

        \draw (-2.5,1.45) to [bend right=45] (0.0,3.0); 
        \draw (-2.5,-1.45) to [bend left=45] (0,-3); 



        \begin{scope}
          \fill[gray!20,opacity=0.7]
            (-2.48,-1.45)
            arc (270:90:-1.75 and 1.45)
            -- cycle;
        \end{scope}


        \coordinate (first_electron_in_S) at (-2.5,-0.52);
        \coordinate (new_first_electron_in_S) at ($(first_electron_in_S)!0.00!(0.6,-1.85)$);
        \coordinate (end) at (-3.5,-0.52);

        \draw[thick, double, double distance=2.5pt, decoration={markings,
                mark=at position 0.50 with {\arrow[scale=0.5,line width=2.5pt]{<}}},
                postaction={decorate}] (new_first_electron_in_S) -- (end)
                node[midway, xshift=-4mm, yshift=-1.5mm] {};

        \coordinate (second_electron_start) at (-2.5,0.50);
        \coordinate (new_second_electron_start) at ($(second_electron_start)!0.00!(0.6,1.55)$);
        \coordinate (end) at (-3.5,0.50);

        \draw[thick, double, double distance=2.5pt, decoration={markings,
                mark=at position 0.5 with {\arrow[scale=0.5,line width=2.5pt]{>}}},
                postaction={decorate}] (new_second_electron_start) -- (end)
                node[midway, xshift=-4mm, yshift=1.5mm] {};

        \coordinate (second_electron_start_in_S) at (-2.5,0.56);
        \coordinate (new_second_electron_start_in_S) at ($(second_electron_start_in_S)!0.00!(0.6,1.55)$);
        \coordinate (end) at (-0.5,1.3);

        \draw[thick, blue, decoration={markings,
                mark=at position 0.55 with {\arrow[scale=0.6,line width=1.5pt]{<}}},
                postaction={decorate}] (new_second_electron_start_in_S) -- (end)
                node[midway, above,  xshift=-0.5mm, yshift=0.5mm] {\scriptsize $e_{k_{1}}^{\sigma}$};


        \coordinate (second_hole_start) at (-2.6,0.4);
        \coordinate (new_second_hole_start) at ($(second_hole_start)!0.03!(0.6,1.55)$);
        \coordinate (end) at (-0.48,1.17);

        \draw[thick, red, decoration={markings,
                mark=at position 0.5 with {\arrow[scale=0.6,line width=1.5pt]{>}}},
                postaction={decorate}] (new_second_hole_start) -- (end)
                node[midway, below, xshift=0.5mm, yshift=-0.5mm] {\scriptsize $h_{-k_{1}}^{-\sigma}$};

        \coordinate (first_electron_start) at (-2.5,-0.58);
        \coordinate (new_first_electron_start) at ($(first_electron_start)!0.00!(0.6,-1.85)$);
        \coordinate (end) at (-0.5,-1.3);

        \draw[thick, blue, decoration={markings,
                mark=at position 0.55 with {\arrow[scale=0.6,line width=1.5pt]{>}}},
                postaction={decorate}] (new_first_electron_start) -- (end)
                node[midway, below, xshift=-0.5mm, yshift=-0.5mm] {\scriptsize $e_{k_{2}}^{\sigma}$};

        \coordinate (first_hole_start) at (-2.5,-0.46);
        \coordinate (new_first_hole_start) at ($(first_hole_start)!0.00!(0.6,-1.75)$);
        \coordinate (end) at (-0.48,-1.17);

        \draw[thick, red, decoration={markings,
                mark=at position 0.5 with {\arrow[scale=0.6,line width=1.5pt]{<}}},
                postaction={decorate}] (new_first_hole_start) -- (end)
                node[midway, above, xshift=-0.5mm, yshift=0.5mm] {\scriptsize $h_{-k_{2}}^{-\sigma}$};

    \node (flux) at (0.0,0) {\textbf{N}};


      \end{tikzpicture}%
    }

    \caption{}
    \label{fig:double_andreev_reflection_diagram} 
  \end{subfigure}

    \vspace{1.0ex}
    

  \caption{(a) Schematic diagram of a normal metal interferometer threaded by an AB flux $\Phi$, coupled to a superconductor. Andreev bound states are shown at the SN interface. (b) For weak coupling to the superconductor, Andreev scattering in the proximity area (gray) leads to conversion of an electron (blue) to a hole (red), adding a Cooper pair (double black line) to the condensate, as well as the inverse process. 
  }
  \label{fig:schematic_andreev_reflection_diagram} 
\end{figure}

\section{Solid-State DSW Experiment}
\label{sec:solid_state_dsw_experiment}
We begin with a brief description and analysis of the original DSW thought experiment. Alice prepares a 
particle in a superposition of spatially separated states
outside a black hole and Bob is inside the black hole \cite{danielsonBlackHolesDecohere2022}. The superimposed matter creates superimposed electromagnetic and gravitational 
fields which travel into the black hole where Bob is able to measure the fields and learn about the superposition. This disturbs the superposition but by causality, Bob's actions cannot have an effect on the superposition. Since he is, in principle, able to make such a measurement, the resolution of this paradox is that Alice's state must disturb itself. 

This thought experiment implies that from Bob's point of view, decoherence is due to the event horizon (or more generally, a Killing horizon \cite{danielsonKillingHorizonsDecohere2023}) harvesting the which-path information of Alice's interferometer \cite{danielsonBlackHolesDecohere2022,grallaDecoherenceHorizonsGeneral2024}. This relativistic point of view demonstrates that the spacetime curvature plays a key role in decoherence. However, there exists a complementary \textit{local} description from Alice's point of view \cite{danielsonLocalDescriptionDecoherence2025} where decoherence is mediated by 
low frequency Hawking radiation \cite{hawkingParticleCreationBlack1975,hawkingBreakdownPredictabilityGravitational1976}. This decoherence mechanism, based as it is on an analysis of quantum field theory in curved spacetime
\cite{danielsonBlackHolesDecohere2022,danielsonKillingHorizonsDecohere2023,grallaDecoherenceHorizonsGeneral2024,danielsonLocalDescriptionDecoherence2025}, is agnostic with respect to the quantum history of the black hole.  It should be equally applicable to young black holes, where the scenario described above of Bob eavesdropping from within the event horizon to obtain which-path information about Alice's interferometer holds, as it is to late-stage black holes which are theorized to act as quantum information mirrors \cite{horowitzBlackHoleFinal2004,haydenBlackHolesMirrors2007,lloydUnitarityBlackHole2014} once they have radiated away half their initial entropy.  In the latter scenario, which-path information incident on the event horizon would be perfectly reflected after a very short time and would propagate out to infinity where it is recorded by the environment (acting as Bob), again causing decoherence.

This alternative scenario is interesting from the lens of analogue studies because superconductors have been noted to be analogues of late-stage black holes \cite{manikandanAndreevReflectionsQuantum2017,manikandanBlackHolesAndreev2020} in SN 
systems \cite{mcmillanTheorySuperconductorNormalMetalInterfaces1968,blonderTransitionMetallicTunneling1982} with the normal metal being the analogue of the exterior of the black hole. In addition, Andreev reflection \cite{andreevThermalConductivityIntermediate1964, pannetierAndreevReflectionProximity2000,beenakkerSpecularAndreevReflection2006,beenakkerColloquiumAndreevReflection2008,houDoubleAndreevReflections2017}, the process wherein an electron (hole) mode in the normal metal is retro-reflected as a hole (electron), with the addition (subtraction) of a Cooper pair in the condensate, has been noted as the analogue of Hawking radiation \cite{jacobsonOriginOutgoingBlack1996}. In this process, the information of the incident electron (hole) is reflected and encoded in the outgoing hole (electron) on a short timescale ${\cal O}(\hbar/\abs{\Delta})$ \cite{averinTimedependentAndreevReflection2020},  where $\Delta(\mathbf{x})=\Delta_{0} e^{i\phi_{S}(\mathbf{x})}$ is the superconducting gap with superconducting phase $\phi_{S}(\mathbf{x})$. An observer in the normal metal unaware of the entanglement of up and down spin electrons induced at the SN interface would interpret the Andreev reflected quasiparticles as arising from thermionic emission \cite{manikandanBlackHolesAndreev2020} from the interface with a temperature \cite{puspusEntanglementSpectrumNumber2014} approximately equal to the critical temperature $T_c$ of the superconductor, which satisfies $2\abs{\Delta}\simeq 3.5 k_B T_c$. Thus $T_c$ plays the role of the Bekenstein-Hawking temperature $T_H$, and the Andreev timescale is consistent (up to a logarithmic factor) with the estimated timescale of late-stage black hole information retrieval \cite{haydenBlackHolesMirrors2007}.

This conceptual one-to-one mapping naturally leads us to consider Alice's point of view and explore the solid-state DSW (SS-DSW) decoherence analogue. Thus, we present the SS-DSW thought experiment: Consider a SN system with a BCS superconductor \cite{bardeenTheorySuperconductivity1957}, which is kept at a temperature $T \ll \abs{\Delta}/k_{B}$. Alice performs a scattering experiment using an Aharonov-Bohm interferometer \cite{aharonovSignificanceElectromagneticPotentials1959,webbObservationAharonovBohmOscillations1985,olariuQuantumEffectsElectromagnetic1985} in N (see Fig.~\ref{fig:schematic_andreev_reflection_diagram}). An electron wave incident from a source lead is coherently split between the two arms of the interferometer, one of which is coupled to the SN interface. The outgoing transmission is measured at a drain lead through its flux-dependent interference contrast. For subgap energies $E\ll \abs{\Delta(\mathbf{x})}$, the branch coupled to S can undergo Andreev conversion, while a second Andreev conversion results in a phase-incoherent transmission into the drain. Bob can collect which-path information by observing the retro-reflected wave packets \cite{benistantDirectObservationAndreev1983} propagating out to infinity in the source and drain electrodes. This is not a false loss of coherence \cite{unruhFalseLossCoherence2000} due to a transient entanglement with the environment, since the information encoded in the entangling particles propagating out to infinity is irretrievable. In Alice's normal metal interferometer, electromagnetic fields are screened, so there is no decoherence no matter how fast the paths are separated and combined at the source and drain electrodes (which is determined by the geometry and the Fermi velocity $v_F$).  Instead, Alice attributes the decoherence in her interferometer to the constant low-frequency power spectrum of Andreev ``thermionic emission.'' This is the solid-state DSW decoherence.

\section{Interferometer Model and Decoherence Metrics}
\label{sec:interferometer_model_and_decoherence_metrics}
To rigorously analyze our SS-DSW thought experiment, we first consider a $N$-site tight-binding ring threaded by an AB flux $\Phi$ (Peierls phase $\phi=e\Phi/\hbar$). Coupled directly to the left side of the ring as a perturbation is a BCS superconductor with (uniform) order parameter $\Delta$. 
This effectively creates a (modified) SN system \cite{mcmillanTheorySuperconductorNormalMetalInterfaces1968,blonderTransitionMetallicTunneling1982} (see Fig.~\ref{fig:general_n_ring_superconductor_diagram}). There are two leads, $\Gamma^{I,II}$, in the wide-band limit coupled to the $N$-ring with coupling strength $\gamma^{I,II}$ that act as a source (lead I) and drain (lead II). An electron is inserted into the ring via lead I, traverses either path of the interferometer, then is absorbed into lead II. Depending on the SN coupling strength $t_{\text{AR}}$, the electron can take a detour into S and undergo Andreev reflection at the SN interface, converting into a retro-reflected hole in the process. An uncorrelated hole can then be converted into an electron that is transmitted into the drain electrode, leading to decoherence (see Fig.~\ref{fig:double_andreev_reflection_diagram}).
We set $T=0$ for all our calculations.

The tight-binding Hamiltonian for a given spin species of the central ring of Alice's interferometer is
\begin{align}
    H_{\text{R}}(\phi) &= \sum_{n=0}^{N-1} \left[\epsilon_{0} c_{n}^{\dagger} c_{n} + t_{0} \left(e^{i\phi/N} c_{n}^{\dagger} c_{n+1} + \text{h.c.} \right)\right], \label{ring_ham}
\end{align}
with $\epsilon_{0}$ and $t_{0}$ being the ring's on-site energy and coupling, respectively, and $n+1$ understood modulo $N$. Here, $c_{n}^{\dagger}$ ($c_{n}$) creates (annihilates) a fermionic particle on site $n$ of the ring. Then the retarded Green's function of Alice's interferometer, without coupling to the superconductor, is $\Tilde{G}^{R}(E,\phi) = \left[E - H_{\text{R}}(\phi) + i \left(\Gamma^{I} + \Gamma^{II} \right)/2 \right]^{-1}$ where $\Gamma^{I,II}$ are tunneling-width matrices describing the coupling of the ring to the source $I$ and drain $II$ electrodes, respectively. 

The coupling of the superconductor to the ring can be described by the Andreev self-energy diagram 
\cite{colemanIntroductionManyBodyPhysics2015} shown in Fig.~\ref{fig:self_energy_diag}, with 
\begin{align}
    \Sigma_{\text{AR}}^{R}(E,\phi) &= \sum_{i,j\in \mathcal{S}} t_{\text{AR}}^{2} \Tilde{G}_{ij}^{R}(E,\phi) \ket{i}\bra{j}, \label{andreev_self_energy}
\end{align}
where $\mathcal{S}$ is the set of sites in the ring coupled to the superconductor, and $t_{\text{AR}}=g(0) t_{\rm NS}^2 Y$ is the effective Andreev coupling at the SN interface, where $g(0)$ is the density of states per unit cell in the normal state of the superconductor, $t_{\rm NS}$ is the tight-binding matrix element at the SN interface, and \cite{nakanoSecondquantizationDescriptionAndreev1994}
\begin{equation}
    Y(E)=\frac{\abs{\Delta}}{\sqrt{\abs{\Delta}^{2} -E^2}}\cos^{-1}\left(\sqrt{\frac{\abs{\Delta} -E}{2\abs{\Delta}}}\right).
    \label{eq:Y}
\end{equation}
Then the \textit{full} retarded Green's function for our model is
\begin{align}
    G^{R}(E,\phi) &= \left[E - H_{\text{R}}(\phi) -\Sigma^{R}(E,\phi) \right]^{-1}, \label{superconductor_on_ring_gf}
\end{align}
with the full retarded self-energy
\begin{align}
    \Sigma^{R}(E,\phi) &= -i\left(\Gamma^I + \Gamma^{II}\right)/2  + \Sigma_{\text{AR}}^{R}(E,\phi).
\end{align}

 \begin{figure}
     \centering
    \begin{equation*}
         \Sigma^{R}_{\text{AR}}(E,\phi) ~=~
         \vcenter{\hbox{\begin{tikzpicture}
       \begin{feynman}
         \vertex[dot] (i) {};
         \vertex [right=2.0cm of i, dot] (o) {};
         \diagram*{
           (i) --[double,double distance=0.5ex,decoration={markings, mark=at position 0.5 with {\arrow[scale=0.5, line width=2.5pt]{>}}},
                    postaction={decorate}] (o)      
         };
       \end{feynman}
       \draw[decoration={markings, mark=at position 0.5 with {\arrow[scale=0.5,line width=2.5pt]{<}}},
             postaction={decorate}]
         (i) to[out=90,in=90,looseness=1.5] (o);
     \end{tikzpicture}}} 
     ~\equiv~
     \hbox{\begin{tikzpicture}
   \begin{feynman}
     \vertex (i1);
     \vertex [right=0.5cm of i1] (i2);
     \vertex [right=0.5cm of i2] (i3);
     \vertex [right=0.5cm of i3] (i4);
     \diagram*{
       (i1) -- [insertion=0.5] (i2) -- [anti fermion, edge label=$-k$] (i3) -- [insertion=0.5] (i4)
     };
   \end{feynman}
 \end{tikzpicture}}
     \end{equation*}
 \caption[Retarded Andreev self-energy $\Sigma^{R}_{\text{AR}}$ diagram.]{
 Andreev self-energy $\Sigma^{R}_{\text{AR}}$ diagram \cite{colemanIntroductionManyBodyPhysics2015}. The double line in the central figure represents the Cooper pair and the two vertices (crosses) in the figure on the right are the Andreev coefficients \cite{nakanoSecondquantizationDescriptionAndreev1994} $t_{\text{AR}}=g(0) t_{\rm NS}^2 Y$.}
 \label{fig:self_energy_diag}
 \end{figure}

To quantify decoherence, we compute the transmission function, contrast, and local density of states (LDOS) of Alice's interferometer
\begin{align}
    T(E,\phi) &= \text{Tr}\left[\Gamma^{I} G^{R}(E,\phi) \Gamma^{II} G^{A}(E,\phi) \right], \label{transmission_formula} \\
    C(E,\phi) &= \frac{2\abs{T(E,\phi) - T(E,0)}}{T(E,\phi) + T(E,0)}, \label{contrast_formula}\\
    \rho(x,E,\phi) &= \braket{x| A(E,\phi) |x}, \label{ldos_formula}
\end{align}
respectively, where $G^{A}=\left[G^{R} \right]^{\dagger}$ is the advanced Green's function, 
\begin{align}
    A(E,\phi) = \frac{1}{2\pi i}\left[G^{A}(E,\phi) - G^{R}(E,\phi) \right],
\end{align}
is the spectral function, and $x$ is one of the sites in the ring. These quantities are evaluated for Alice's linear-response transport measurements at $E=E_F=0$, which is equal to the center of the superconducting gap in equilibrium, and can be varied relative to the energy levels in the interferometer by an appropriate gate voltage which shifts $\varepsilon_0$.  The Andreev coupling function in Eq.~\eqref{eq:Y} thus takes the value $Y(0)=\pi/4$.

The transmission function gives the probability that an injected electron
reaches the drain, summed over incoming and outgoing channels \cite{dattaElectronicTransportMesoscopic1997}. The contrast describes how strongly $T(E,\phi)$ oscillates with the AB flux $\Phi$, and is a direct measure of coherence. The LDOS describes how much spectral weight is available at site $x$ \cite{stefanucciNonequilibriumManyBodyTheory2025}. 

All simulations are carried out at absolute zero temperature (so there is no decoherence due real thermal fluctuations) with 
parameters: $N=100$, $\phi=\pi$ (corresponding to one-half magnetic flux quantum), $E_F=0$, 
$t_{0}=t_{x}=t_{y}=t_{x}^{\prime}=-1$, $\abs{\Delta}=1$, $\gamma 
=0.2$, rank=20 $\Gamma^{I,II}$ matrices, and $\mathcal{S}=\{20,\ldots, 29\}$ (so $M_{y}=\abs{\mathcal{S}}=10$ superconducting sites).

\begin{figure}[t]
  \centering

  \begin{subfigure}{\columnwidth}\centering
    \includegraphics[width=\linewidth]{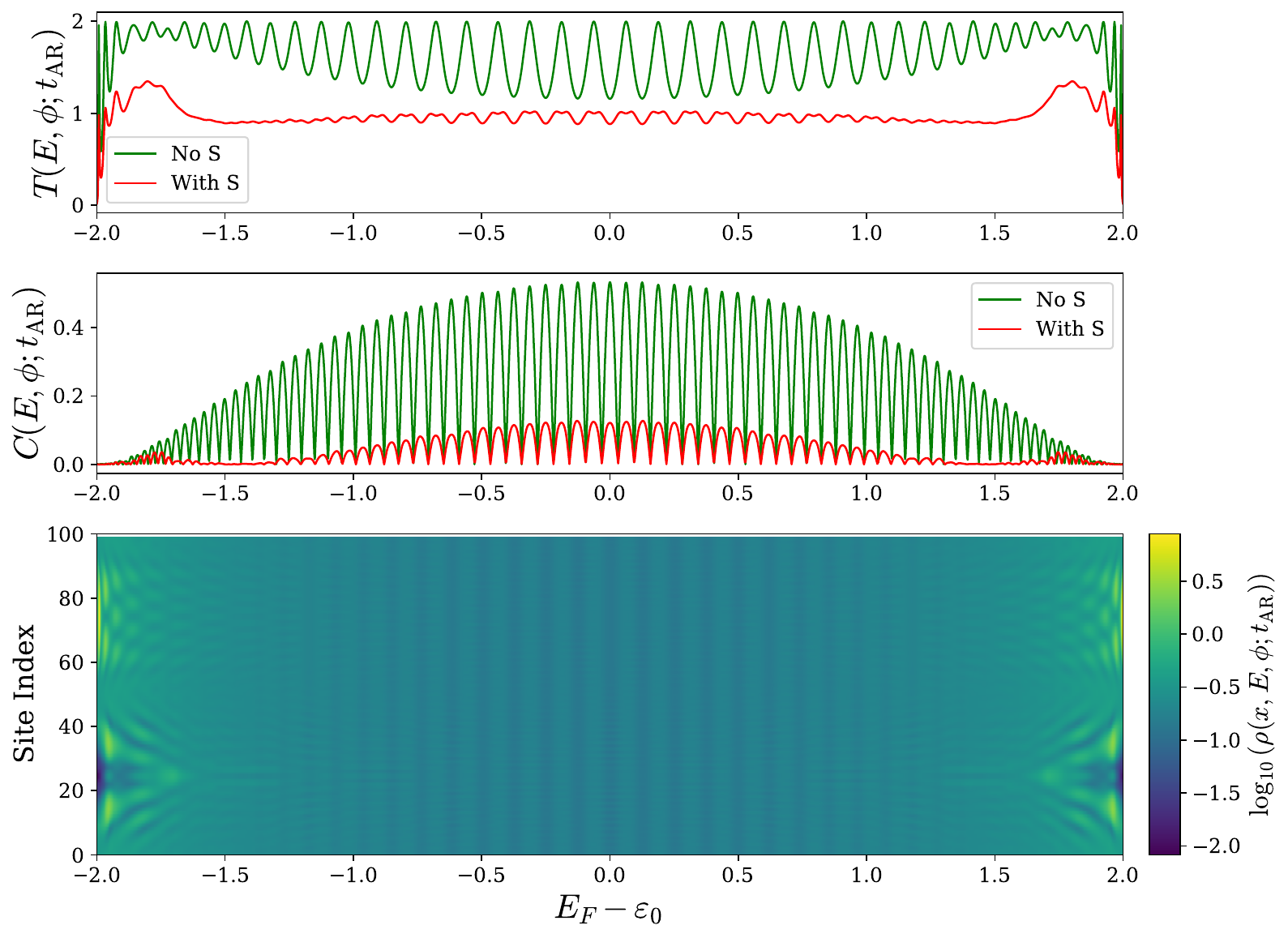}
    \caption{\centering $t_{\text{AR}}=0.2$}
    \label{fig:weak_coupling_tAR0.200TxCxLDOS_phi00.000_phi13.142_Mx0_My10_N100_rank20_G0.200_E2001_logLDOS1}
  \end{subfigure}
  
  \begin{subfigure}{\columnwidth}\centering
    \includegraphics[width=\linewidth]{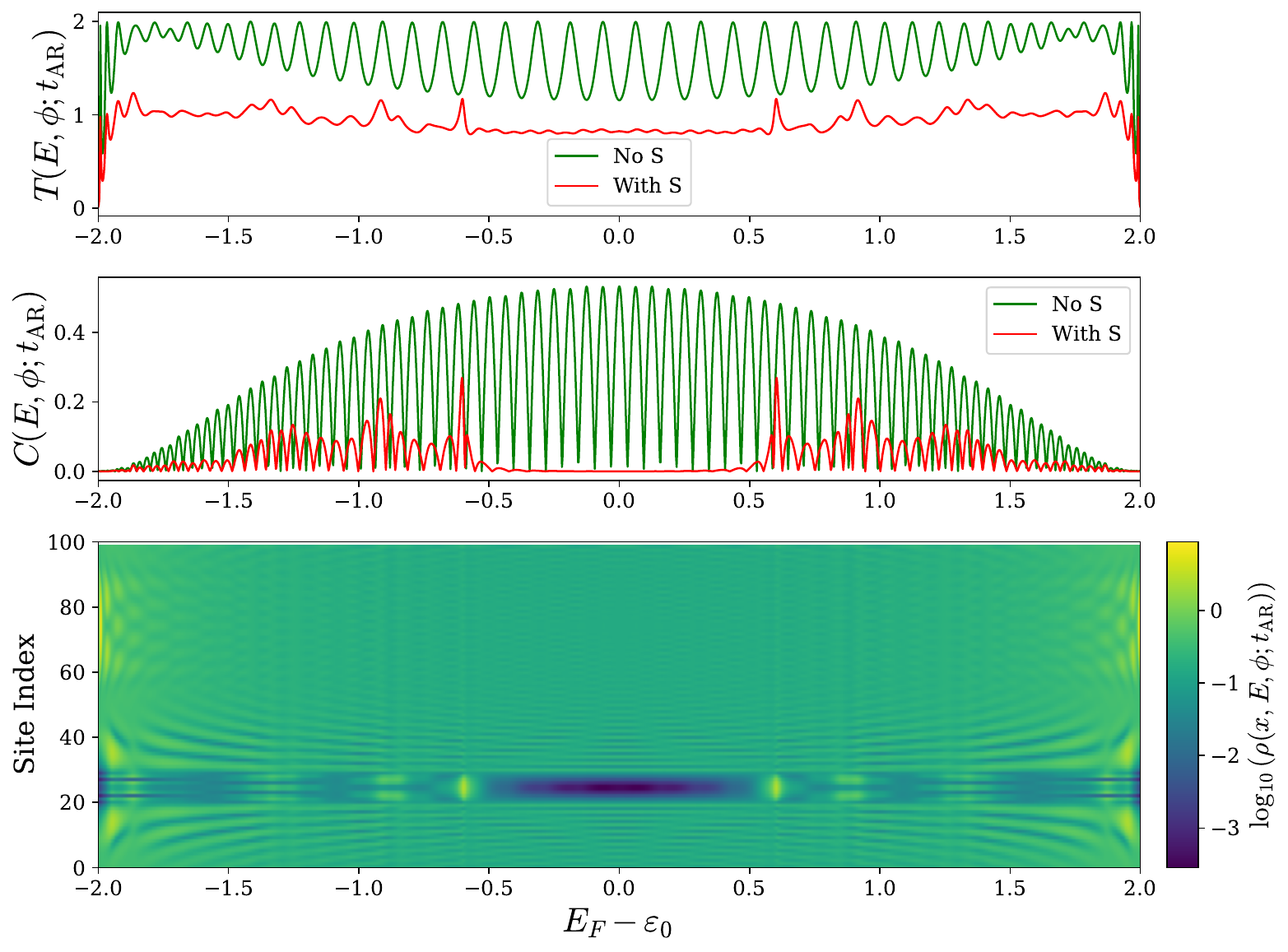}
    \caption{\centering $t_{\text{AR}}=2.5$}
    \label{fig:strong_coupling_tAR2.500TxCxLDOS_phi00.000_phi13.142_Mx0_My10_N100_rank20_G0.200_E2001_logLDOS1}
  \end{subfigure}

  \caption{Transmission $T(E,\phi; t_{\text{AR}})$, contrast $C(E,\phi; t_{\text{AR}})$, and LDOS $\rho(x,E,\phi; t_{\text{AR}})$ at $E=E_F$ vs $E_F-\varepsilon_0$ for two values of $t_{\text{AR}}$, computed at {\it absolute zero temperature}. We observe decoherence in Fig.~\ref{fig:weak_coupling_tAR0.200TxCxLDOS_phi00.000_phi13.142_Mx0_My10_N100_rank20_G0.200_E2001_logLDOS1} and the reemergence of coherence in Fig.~\ref{fig:strong_coupling_tAR2.500TxCxLDOS_phi00.000_phi13.142_Mx0_My10_N100_rank20_G0.200_E2001_logLDOS1}.  See ancillary file for a video illustrating the evolution of $T$, $C$, and $\rho$ as a function of $t_{\text{AR}}$.}
  \label{fig:TxCxLDOS_triptych_row}
\end{figure}

\begin{figure}[t]
  \centering

    \begin{subfigure}{\columnwidth}\centering
    \includegraphics[width=\linewidth]{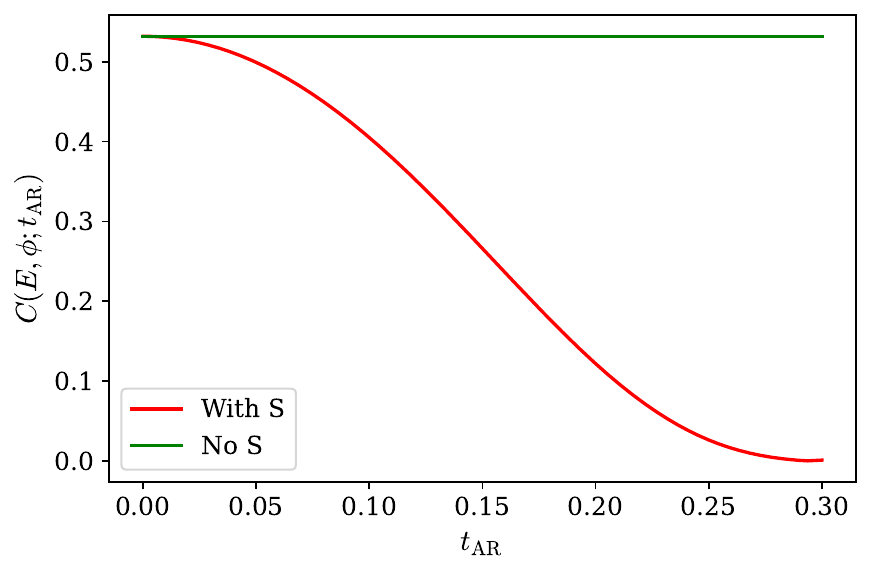}
    \caption{\centering}
    \label{fig:C_vs_tAR_weak_coupling_regime}
  \end{subfigure} \hfill

   \begin{subfigure}{\columnwidth}\centering
    \includegraphics[width=\linewidth]{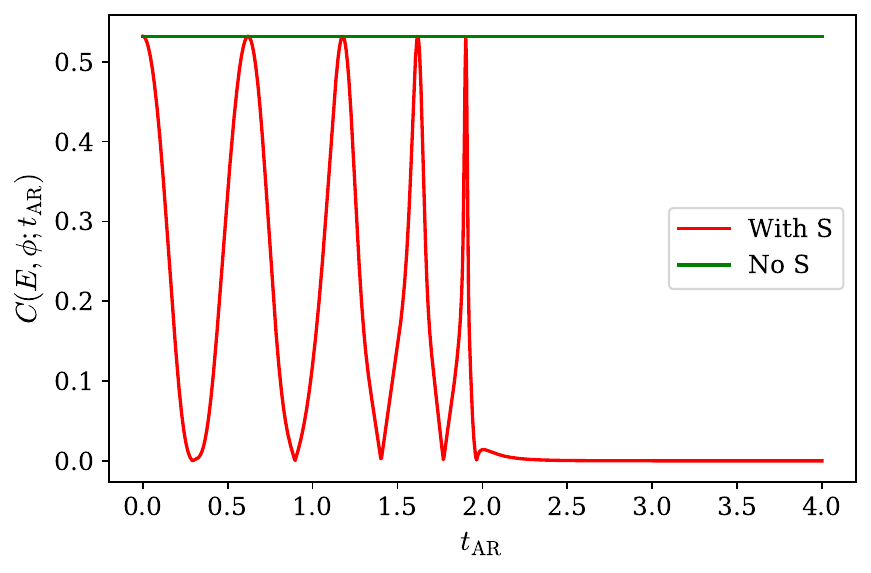}
    \caption{\centering}
    \label{fig:C_vs_tAR_full_tAR_regime}
  \end{subfigure} 

  \caption{Contrast $C(E,\phi ; t_{\text{AR}})$ as a function of $t_{\text{AR}}$, computed at {\it absolute zero temperature}. (a) \textit{Weak coupling regime}. We see contrast suppression in the weak coupling regime until $t_{\text{AR}}=0.3$. (b) \textit{Full regime}. We see the oscillatory behavior of the reemergence of coherence until the final contrast suppression in the strong coupling regime. 
  }
  \label{fig:contrast_vs_tAR}
\end{figure}

\section{Decoherence and Reemergence of Coherence}
\label{sec:decoherence_and_reemergence_of_coherence}
The transmission function $T$, contrast $C$, and LDOS $\rho$ are plotted in the limits of weak, intermediate, and strong coupling to the superconductor in Fig.~\ref{fig:TxCxLDOS_triptych_row}, where the green curves show the transmission and contrast without the coupling to the superconductor, and the red curves include the superconductor.
In the \textit{weak} coupling regime $0 \leq t_{\text{AR}} \leq 0.3$,  the transmission and contrast are suppressed due to Andreev reflection, demonstrating decoherence near the SN interface (see Figs.~\ref{fig:weak_coupling_tAR0.200TxCxLDOS_phi00.000_phi13.142_Mx0_My10_N100_rank20_G0.200_E2001_logLDOS1}
and \ref{fig:C_vs_tAR_weak_coupling_regime}). In this regime, incoherent transmission due to the lowest (2nd) order Andreev scattering processes dominates, thus reducing the contrast. 

Interestingly, in the \textit{intermediate} coupling regime $0.3 \leq t_{\text{AR}} \leq 2$, higher-order Andreev processes start to dominate, resulting in the reemergence of coherence near the SN interface (see Fig.~\ref{fig:C_vs_tAR_full_tAR_regime}). This is because two successive AR close the detour into a coherent return channel, so the which-path leakage becomes a reversible virtual process, yielding the return of AB contrast. This reemergence of coherence occurs even with disorder in our system, demonstrating that this phenomenon is rather robust (see App.~\ref{app:disorder_comparison}). In addition, it does not occur when the superconductor is replaced with a normal reservoir due to the fact that mode-conversion processes, such as Andreev reflection, are responsible for this phenomenon (see App.~\ref{app:normal_reservoir_comparison}).  We note that a reemergence of coherence has also been found in other black hole analogue models \cite{anglinDecoherenceRecoherenceAnalogue1995,linEntanglementRecoherenceInformation2008}. 
In the gravitational analogue problem, we speculate that the reemergence of coherence would correspond to coherent transmission mediated by virtual Hawking radiation. Such processes may be particularly relevant for near-extremal black holes, which are theorized to possess critical quantum fluctuations \cite{maldacenaConformalSymmetryIts2016,maulikQuantumcorrectedHawkingRadiation2026}.

The \textit{strong} coupling regime $t_{\text{AR}} \geq 2$ shows pronounced subgap spectral features localized along the SN interface, as seen in the LDOS plot of Fig.~\ref{fig:strong_coupling_tAR2.500TxCxLDOS_phi00.000_phi13.142_Mx0_My10_N100_rank20_G0.200_E2001_logLDOS1} and schematically in Fig.~\ref{fig:general_n_ring_superconductor_diagram}. These standing waves are associated with resonant peaks in the transmission and contrast, corresponding to proximity-induced Andreev bound states which are coherent electron-hole resonances generated by repeated Andreev conversion \cite{andreevElectronSpectrumIntermediate1966,kulikMacroscopicQuantizationProximity1969,beenakkerQuantumTransportSemiconductorsuperconductor1992,marmorkosThreeSignaturesPhasecoherent1993,lenssenExperimentalSignaturePhasecoherent1998,bretheauExcitingAndreevPairs2013,saulsAndreevBoundStates2018,haysDirectMicrowaveMeasurement2018}.  In this regime, the real part (Lamb shift) of the Andreev self-energy $\Sigma_{\text{AR}}^{R}$ pushes the sites along the SN interface away from the band center, leading to reduced AB contrast in the transmission (see LDOS plot of Fig.~\ref{fig:strong_coupling_tAR2.500TxCxLDOS_phi00.000_phi13.142_Mx0_My10_N100_rank20_G0.200_E2001_logLDOS1}). Further increases in  coupling lead to $T(E,\phi)$ and $C(E,\phi)$ converging to 1 and 0, respectively, signaling that the sites coupled to S are essentially blocked by a very large Lamb shift.

\section{Mathematical Analogy to Gravitation}
\label{sec:mathematical_analogy_to_gravitation}
Recall that the Unruh effect \cite{fullingNonuniquenessCanonicalField1973,daviesScalarProductionSchwarzschild1975,unruhNotesBlackholeEvaporation1976} is formulated via a Bogoliubov transformation between inertial (Minkowski) and uniformly accelerated (Rindler) mode operators, under which the Minkowski vacuum becomes an entangled state across the causally disconnected left and right Rindler wedges \cite{crispinoUnruhEffectIts2008,uedaEntanglementVacuumLeft2021a}. For the Dirac (fermionic) case, this structure is obtained by analytic continuation of positive-frequency $\omega$ mode solutions across the Rindler horizon, leading to mode relations that explicitly contain the Unruh factor $e^{-\pi\omega c/a}$, where $a$ is the proper acceleration \cite{soffelDiracParticlesRindler1980,uedaEntanglementVacuumLeft2021a}. Schematically, the Minkowski vacuum $\ket{0^{\text{M}}}$ may be written as an entangled superposition of excitations built on the Rindler product vacuum $\ket{0_{\text{I}} 0_{\text{II}}}$ of the wedge-I $\ket{0_{\text{I}}}$ and wedge-II $\ket{0_{\text{II}}}$ vacua (see Ref.~\cite[Eq.~(160)]{uedaEntanglementVacuumLeft2021a})
\begin{align}
    \ket{0^{\text{M}}} &\propto \prod_{\omega,\mathbf{k}_{\perp},n} \left(1 + e^{-\pi\omega c/a} c_{\omega,\mathbf{k}_{\perp}}^{\text{I},n \dagger} d_{\omega,-\mathbf{k}_{\perp}}^{\text{II}, \bar{n} \dagger} \right) \notag \\
    &\times\left(1 - e^{-\pi\omega c/a} c_{\omega,\mathbf{k}_{\perp}}^{\text{II},n \dagger} d_{\omega,-\mathbf{k}_{\perp}}^{\text{I},\bar{n} \dagger} \right) \ket{0_{\text{I}} 0_{\text{II}}}, \label{rindler_vacuum_state}
\end{align}
where $c_{\omega,\mathbf{k}_{\perp}}^{\text{i},n \dagger}$ ($c_{\omega,\mathbf{k}_{\perp}}^{\text{i},n }$) creates (annihilates) a particle with energy $\omega$, spin $n$, and transverse momentum $\mathbf{k}_{\perp}$ in wedge i of Minkowski space and $d_{\omega,\mathbf{k}_{\perp}}^{\text{i},n \dagger}$ ($d_{\omega,\mathbf{k}_{\perp}}^{\text{i},n }$) creates (annihilates) an antiparticle with energy $\omega$, spin $n$, and transverse momentum $-\mathbf{k}_{\perp}$ in wedge i of Minkowski space. A uniformly accelerated observer restricted to a single wedge therefore describes the Minkowski vacuum by a thermal reduced density operator after tracing over the inaccessible wedge \cite{soffelDiracParticlesRindler1980,uedaEntanglementVacuumLeft2021a}.

Strikingly, the same mathematical structure appears in the superconducting case: the BCS ground state $\ket{\text{BCS}}$ is the Bogoliubov vacuum of Bogoliubov-de Gennes (BdG) quasiparticles, but when written in terms of the underlying electron operators it is a paired Gaussian state with a Schmidt-like product structure under an appropriate partition \cite{bardeenTheorySuperconductivity1957,puspusEntanglementSpectrumNumber2014}
\begin{align}
    \ket{\text{BCS}} &= \prod_{\mathbf{k}} \left(u_{\mathbf{k}} + v_{\mathbf{k}} c_{\mathbf{k}\uparrow}^{\dagger}c_{-\mathbf{k}\downarrow}^{\dagger} \right) \ket{0}. \label{bcs_vacuum_state}
\end{align}
A key distinction is that the Nambu doubling reorganizes a single electronic Fock space into particle-hole sectors relative to the Fermi energy $E_{F}$, whereas in relativistic Dirac theory, particles/antiparticles are associated with distinct creation/annihilation operators restricted to positive energies. This motivates a unified viewpoint in which ``particle content'' depends on the chosen mode decomposition, while reduced descriptions obtained by tracing out inaccessible sectors acquire effective thermality.

Importantly, the distribution 
produced by the Bogoliubov mixing in superconductors is generally not exactly thermal. At absolute zero temperature, tracing over one spin sector of a paired Gaussian ground state yields a reduced state whose entanglement spectrum is described by a generalized Gibbs ensemble with a mode-dependent reciprocal temperature $\beta_{e}(\xi)$ \cite{puspusEntanglementSpectrumNumber2014}. Near the Fermi surface, this can be well approximated by a grand canonical ensemble with $\beta = 2/\abs{\Delta}$. The associated entanglement entropy obeys an ``area law'' scaling, $S \propto \pi g(0) \abs{\Delta}$, and is directly tied to number fluctuations \cite{puspusEntanglementSpectrumNumber2014}.  

The connection between the SN model and the Dirac field theory in Rindler space is not a coincidence \cite{volovikSuperfluidAnalogiesCosmological2001,volovikUniverseHeliumDroplet2009}. At subgap energies near the Fermi surface, $\abs{E-E_{F}} < \abs{\Delta}$, Alice's interferometer is naturally described by the continuum Nambu spinor BdG theory \cite{bogolyubovTheorySuperfluidity1947,gennesSuperconductivityMetalsAlloys2018} rather than the full microscopic lattice model. In the low-energy limit (Andreev approximation \cite{andreevThermalConductivityIntermediate1964}), we can linearize the normal-state dispersion about the Fermi momentum $p_{F}=\hbar k_{F}$ and bundle the electron $u$ and hole $v$ amplitudes into a Nambu spinor $\Psi=\left(u,v \right)^{T}$ \cite{nambuQuasiParticlesGaugeInvariance1960}. This effectively reduces the Andreev dynamics to a first-order Dirac-like equation 
\begin{align}
    i\hbar\partial_{t}\Psi &= \left[-i\hbar \tau_{3}\mathbf{v}_{F} \cdot \nabla_{\perp}   +\vec{\Delta}(\mathbf{x},t) \cdot \vec{\tau} \right] \Psi,
\end{align}
where the pairing field $\vec{\Delta}(\mathbf{x}) = \Delta_{0}(\mathbf{x})\left(\cos\phi_{S}(\mathbf{x}), -\sin\phi_{S}(\mathbf{x}),0 \right)$ acts as a mass-like term, $\vec{\tau}=\left(\tau_{1}, \tau_{2}, \tau_{3}\right)$ are the $2\times 2$ Nambu matrices, $\mathbf{v}_{F}(\mathbf{k})$ is the Fermi velocity vector, $\nabla_{\perp} \equiv \left(\partial_{x},\partial_{y} \right)$, and the electron-hole conversion is encoded by the off-diagonal Bogoliubov mixing \cite{volovikSuperconductivityLinesGAP1993,volovikSuperfluidAnalogiesCosmological2001,volovikUniverseHeliumDroplet2009} (see Supplementary Material \cite{SupplementalMaterialsungDecoherenceReemergenceCoherence2026} for derivation). In this sense, the low-energy sector of the SN system has the same spinor-field structure as the Dirac field theory in Rindler space, with the condensed matter doubling being a Nambu particle-hole doubling relative to $E_{F}$ rather than the particle-antiparticle doubling of relativistic quantum field theory.

\section{Distance Dependence of Decoherence Rate}
\label{sec:distance_dependence_of_decoherence_rate}
To calculate the distance dependence of the SS-DSW decoherence, we introduce a 2D normal metal spacer between the $N$-ring and S with $M_{x}$ horizontal and $M_{y}$ vertical sites, giving a total of $M_{x}M_{y}$ sites (
details given in the Supplementary Material \cite{SupplementalMaterialsungDecoherenceReemergenceCoherence2026}).

\begin{figure}[t]
  \centering
    \includegraphics[width=\linewidth]{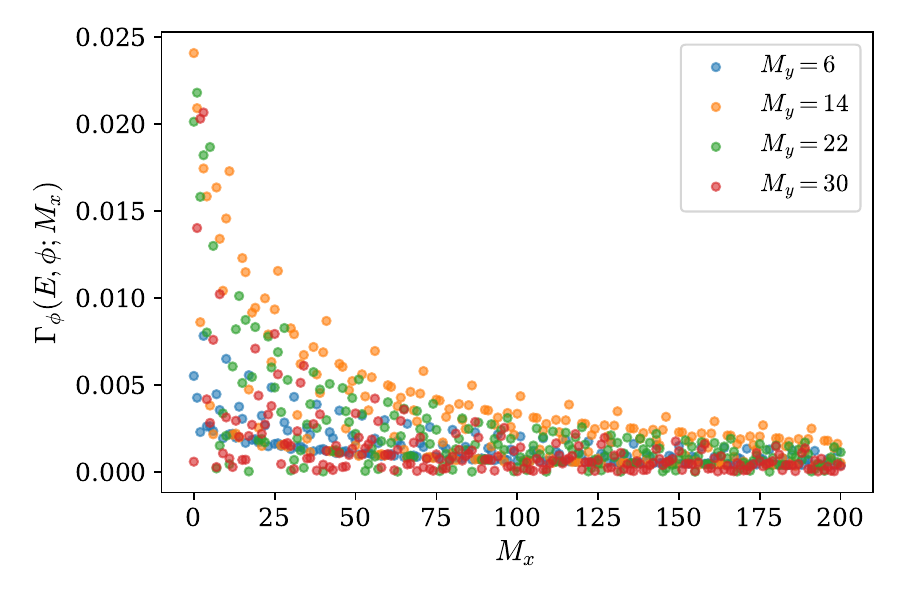}

  \caption{Dephasing $\Gamma_{\phi}(E,\phi; M_{x})$ vs horizontal sites $M_{x}$ for weak coupling $t_{\text{AR}}=0.2$ and $M_{y}\in\{6,14,22,30\}$, calculated at absolute zero temperature. The dephasing exhibits an approximate $1/M_{x}$ ($1/r$) trend. We use the fixed preset parameters and $t_{\text{AR}}=0.2$. 
  }
  \label{fig:tau_inv_vs_Mx_multiMy}
\end{figure}

To quantify the spatial dependence of the decoherence rate, we utilize the transport-weighted  \cite{staffordLocalTemperatureInteracting2016} dephasing rate 
\begin{align}
    \frac{\Gamma_{\phi}(E,\phi)}{\hbar} &= -\frac{2 \text{Im}\text{Tr}\left[\Sigma^R_{\text{AR}}(E,\phi) G^{R}(E,\phi) \Gamma^{I} G^{A}(E,\phi) \right]}{\hbar \, \text{Tr}\left[G^{R}(E,\phi) \Gamma^{I} G^{A}(E,\phi) \right]}. \label{transport_weighted_dephasing}
\end{align}
We see in Fig.~\eqref{fig:tau_inv_vs_Mx_multiMy} that for multiple S sites on the spacer, the dephasing $\Gamma_{\phi}(E,\phi;M_{x})$ as a function of $M_{x}$ exhibits an approximate $1/M_{x}$ ($1/r$) trend. The spread among the different $M_{y}$ curves is a quantum-size effect wherein the discrete waveguide modes of the mesoscopic spacer modulate the coupling of the superconductor to the ring, producing geometry-dependent oscillations around the same decaying envelope. We see that each $M_{y}$ curve exhibits an approximate $1/M_{x}$ ($1/r$) trend, in qualitative agreement with the DSW decoherence scaling of $1/r^{3}$ \cite{danielsonBlackHolesDecohere2022,danielsonKillingHorizonsDecohere2023,grallaDecoherenceHorizonsGeneral2024,danielsonLocalDescriptionDecoherence2025}. We note that the original DSW decoherence scaling arises in a three-dimensional geometry, whereas our model is effectively one-dimensional, so the corresponding distance dependence is naturally weaker. 

In the SN structure, $\Delta$ plays the role of an effective horizon for subgap modes: there are no propagating quasiparticle excitations in S for $\abs{E} < \abs{\Delta}$ at $T=0$, while the influence of S on N is confined to a proximity layer of thickness $\xi \sim \hbar v_{F} / \abs{\Delta}$ (the superconducting coherence length), which shrinks as $\abs{\Delta} \to \infty$ \cite{blonderTransitionMetallicTunneling1982}. In this limit, a spin-diagonal probe restricted to the normal region at distances $x \gg \xi$ effectively accesses only the reduced state of N, while subgap transport is governed by Andreev reflection, which implements electron-hole Bogoliubov mixing. In this sense, the proximity region plays the role of an analogue near-horizon (Rindler-like) region, while the superconducting condensate acts as the analogue black hole interior, where infalling matter is maximally entangled with infalling Hawking quanta \cite{haydenBlackHolesMirrors2007}.

\section{Discussion}
\label{sec:discussion}
We have demonstrated a solid-state analogue of the DSW decoherence in a superconductor-normal metal AB interferometer. At weak SN coupling $t_{\text{AR}}$, Andreev processes suppress the interference contrast, providing a condensed matter analogue of horizon-induced decoherence. In an interesting twist, we found a reemergence of coherence at intermediate coupling, which we interpret as due to resonant tunneling through Andreev bound states.

Our results support a unified horizon analogy in which the superconducting gap defines an effective inaccessible sector, Andreev reflection plays the role of Hawking radiation, and the low-energy (near Fermi surface) effective field theory takes on a Dirac-like Nambu spinor form. At an SN interface, Andreev reflection and the proximity effect may be understood as two complementary manifestations of the same interface-induced entangling process: one visible in the quasiparticle scattering channels, the other in the induced superconducting correlations of the background state \cite{nakanoSecondquantizationDescriptionAndreev1994}. Hawking radiation admits a closely related dual description, namely, in terms of emitted quanta produced by horizon-mediated mode conversion \cite{hawkingParticleCreationBlack1975} and in terms of the altered vacuum structure near the horizon \cite{unruhNotesBlackholeEvaporation1976}. In both cases, what appears as particle production from one perspective is inseparable from a reorganization of the underlying ground state from another. 

This solid-state analogue of DSW decoherence should not be conflated with the decoherence induced by thermal fluctuations of a material body with internal degrees of freedom chosen to mimic the effects of a black hole \cite{biggsComparingDecoherenceEffects2024,danielsonLocalDescriptionDecoherence2025}; it occurs even if the SN interferometer is in its quantum mechanical ground state. In that sense, the analogy to horizon-induced decoherence is much closer since the apparent thermal fluctuations in both cases are determined by mode-conversion processes described by analogue Bogoliubov transformations [cf., Eqs.\ \eqref{rindler_vacuum_state}, \eqref{bcs_vacuum_state}]. To this end, we conjecture that any system lacking such mode conversion is not a proper black hole analogue.

Beyond merely theoretical interest, this one-to-one conceptual mapping opens the possibility of leveraging superconducting interferometers as experimental platforms for studying decoherence from horizon-like physics. Given the ubiquity of Hawking-like radiation in various analogue systems, we hypothesize that similar analogue DSW decoherence could exist in other analogue platforms. On a speculative level, our results suggest that the predicted coherence restoration mediated by virtual Hawking radiation may be observable in the vicinity of near-extremal black holes, whose quantum fluctuations are theorized to be critical. SN systems thus represent a promising new analogue platform to study the interplay of quantum information and gravitation.

\section{Note added}
\label{sec:note_added}
After submission of this manuscript, we became aware of a recent work \cite{biggsNotAllBlack2026} claiming that decoherence is indeed suppressed for near extremal black holes due to their large quantum metric fluctuations.

\section{Acknowledgments}
\label{sec:acknowledgments}
We thank Samuel E. Gralla, Morifumi Mizuno, Parth Kumar, Marco A. Jimenez-Valencia, Anand H. Natarajan, and Xingzhou Yu for the helpful comments and discussions in the formulation of this work. 

\section{Data availability}
\label{sec:data_availability}
The data supporting the findings of
this article are openly available \cite{sungDatasetDecoherenceReemergence2026}.


\bibliography{Superconductors_and_Black_Holes}


\onecolumngrid

\begin{center}
    \bfseries End Matter
\end{center}

\twocolumngrid

\setcounter{secnumdepth}{1}


\makeatletter

\let\PRLoriginalsection\section

\let\PRLoriginalhangfromsection\@hangfrom@section

\let\PRLoriginalsectioncntformat\@sectioncntformat

\let\PRLoriginalrunintosection\@runin@to@section

\makeatother


\appendix


\let\PRLoriginalsection\section

\renewcommand{\section}[1]{%
    \PRLoriginalsection{%
        \texorpdfstring{#1}{Appendix \thesection: #1}%
    }%
}

\makeatletter

\def\@runin@to@section#1#2#3{#1#2:\ #3---}

\makeatother

\section{Effect of Disorder on Decoherence\label{app:disorder_comparison}}

We demonstrate that the decoherence and reemergence of coherence exists even in the presence of disorder in the interferometer ring. See Sec.~\ref{sec:disordered_ring} of the Supplementary Material \cite{SupplementalMaterialsungDecoherenceReemergenceCoherence2026} for a more detailed derivation and discussion. 

To include disorder in our model, we add to the AB ring Hamiltonian \eqref{ring_ham} a random potential $V_{n}^{(s)}$ that is independently sampled from a uniform distribution ($V_{n}^{(s)} \sim \mathcal{U}\left(-W\abs{t_{0}},+W\abs{t_{0}} \right)$) where $W \geq 0$ is the dimensionless disorder strength. The superscript $(s)$ labels an independent disorder realization, or seed. Thus, the disordered $N$-site tight-binding AB ring is
\begin{align}
    H_{\text{R}}^{(s)}(\phi) &= \sum_{n=0}^{N-1} \left[\left(\varepsilon_{0} + V_{n}^{(s)} \right) c_{n}^{\dagger}c_{n} \right. \notag \\
    &\left.\quad+ t_{0} \left(e^{i\phi/N} c_{n}^{\dagger} c_{n+1} + \text{h.c.} \right) \right]. \label{disorder_ab_ring_ham}
\end{align}
Then the retarded Green's function of Alice's disordered interferometer \eqref{disorder_ab_ring_ham} is
\begin{align}
    \Tilde{G}^{R(s)}(E,\phi) &= \left[E - H_{\text{R}}^{(s)}(\phi) +i \left(\Gamma^{I} + \Gamma^{II} \right)/2\right]^{-1}. \label{disordered_ab_ring_gf}
\end{align}
The next steps are trivial if we recall that the non disordered transmission \eqref{transmission_formula}, contrast \eqref{contrast_formula}, and LDOS \eqref{ldos_formula} relied on the full retarded Green's function \eqref{superconductor_on_ring_gf} of our model. However, Eq.~\eqref{superconductor_on_ring_gf} relied on the Andreev self-energy \eqref{andreev_self_energy} which in turn relied on the non disordered AB ring's retarded Green's function $\Tilde{G}^{R}(E,\phi) = \left[E - H_{\text{R}}(\phi) + i \left(\Gamma^{I} + \Gamma^{II} \right)/2 \right]^{-1}$. This means that the replacement $\Tilde{G}^{R} \to \Tilde{G}^{R(s)}$ in Eqs.~\eqref{andreev_self_energy} and \eqref{superconductor_on_ring_gf} gives the disordered Andreev self-energy $\Sigma_{\text{AR}}^{R(s)}$ and full retarded Green's function 
\begin{align}
    G^{R(s)}(E,\phi) &= \left[E - H_{\text{R}}^{(s)}(\phi) - \Sigma^{R(s)}(E,\phi)\right]^{-1} \label{full_gf_disorder}
\end{align}
of our model, respectively. This yields the disordered transmission, contrast, and LDOS 
\begin{align}
    &T^{(s)}(E,\phi) = \text{Tr}\left[\Gamma^{I} G^{R(s)}(E,\phi) \Gamma^{II} G^{A(s)}(E,\phi) \right], \\
    &C^{(s)}(E,\phi) = \frac{2 \abs{T^{(s)}(E,\pi) - T^{(s)}(E,0)}}{T^{(s)}(E,\pi) + T^{(s)}(E,0)}, \\
    &\rho^{(s)}(x,E,\phi) = \braket{x| A^{(s)}(E,\phi) |x},
\end{align}
respectively, for an \textit{individual} disorder seed $s$, where $A^{(s)}(E,\phi)$ is the disordered spectral function.

\begin{figure}[t]
  \centering
    \includegraphics[width=\linewidth]{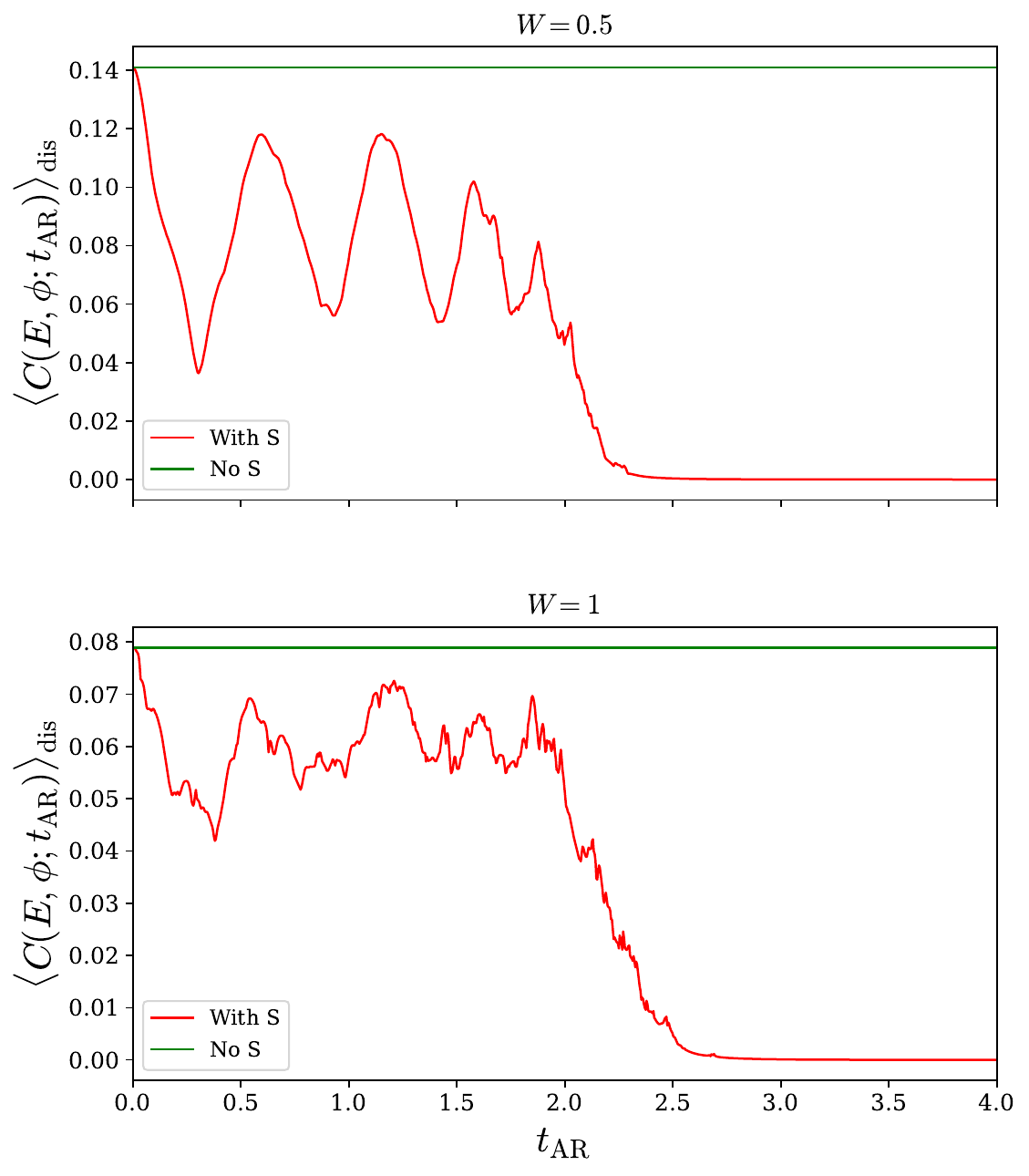}
    \caption{Ensemble averaged contrast $\braket{C(E,\phi; t_{\text{AR}})}_{\text{dis}}$ as a function of $t_{\text{AR}}$ for disorder strength $W=0.5$ and $W=1$ at the band center $E=0$, computed at absolute zero temperature. We observe the same decoherence and reemergence of coherence behavior even with disorder.}
    \label{fig:C_vs_tAR_full_and_zoom_2x2}
\end{figure}

The ensemble averaged transmission, contrast, and LDOS 
\begin{align}
    \braket{T(E,\phi)}_{\text{dis}} &= \frac{1}{N_{s}} \sum_{s=1}^{N_{s}} T^{(s)}(E,\phi), \\
    \braket{C(E,\phi)}_{\text{dis}} &= \frac{1}{N_{s}} \sum_{s=1}^{N_{s}} C^{(s)}(E,\phi), \\
    \braket{\rho(x,E,\phi)}_{\text{dis}} &= \frac{1}{N_{s}} \sum_{s=1}^{N_{s}}  \rho^{(s)}(x,E,\phi), 
\end{align}
respectively, of Alice's disordered interferometer are plotted in Fig.\ \ref{fig:C_vs_tAR_full_and_zoom_2x2}. 
For all of our disorder simulations, we use the same parameters as used in the main text with $N_{s}=100$ seeds.

For weak disorder ($W=0.5$) the mean transmission is reduced by a factor of 2, while for stronger disorder ($W=1.0$) the mean transmission is reduced by a factor of 5, representing diffusive transport, not localization that would render the interferometer an open circuit. (See Sec.~\ref{sec:disordered_ring} of the Supplementary Material \cite{SupplementalMaterialsungDecoherenceReemergenceCoherence2026}.)

    For both disorder strengths, the disordered interferometer reveals the same physics as for the pristine interferometer ring: a monotonic reduction of interference contrast for weak coupling to the superconductor $t_{\text{AR}} \leq 0.3$ (0.4 for stronger disorder), then a reemergence of coherence due to resonant tunneling through Andreev bound states 
    for intermediate coupling, followed by a complete suppression of coherence for sufficiently strong coupling to the superconductor.

    This validates our claim that the predicted analogue quantum gravitational effects should be observable in real samples.

\section{Replacing the Superconductor With a Normal Reservoir}
\label{app:normal_reservoir_comparison}

We demonstrate that the signatures of the reemergence of coherence are not replicated when replacing the BCS superconductor with a normal reservoir. We also state the importance of mode-conversion processes in analogue black holes. See Sec.~\ref{sec:normal_reservoir} of the Supplementary Material \cite{SupplementalMaterialsungDecoherenceReemergenceCoherence2026} for a more detailed derivation and analysis.

Suppose that each ring site $s_a\in\mathcal{S}$ previously coupled to the superconductor is instead connected to an independent semi-infinite normal tight-binding wire, each of which is connected to a common normal fermionic reservoir at infinity. The Hamiltonian of wire $a$ and its coupling to the ring are
\begin{align}
&H_a = \sum_{n=0}^{\infty}\left[\epsilon_w d_{a,n}^{\dagger}d_{a,n} + t_w\left(d_{a,n+1}^{\dagger}d_{a,n} + \text{h.c.}\right)\right], \\
&H_{\text{R}w,a} = \tau_w c_{s_a}^{\dagger}d_{a,0} + \tau_w^{*}d_{a,0}^{\dagger}c_{s_a},
\end{align}
respectively, where $\epsilon_w$ and $t_w$ are the onsite energy and nearest-neighbor hopping of the normal wires, and $\tau_w$ is the ring-wire hopping amplitude. Although the $N_c=\abs{\mathcal{S}}$ wires are independent channels, they are coupled to a single floating normal reservoir, denoted by terminal $p=3$. We analytically integrate out the degrees of freedom of the semi-infinite wires to get the self-energy
\begin{align}
\Sigma_p^R(E) &= \left[\Lambda_p(E)-\frac{i}{2}\Gamma_p(E)\right]\sum_{a=1}^{N_c}\ket{s_a}\bra{s_a},
\end{align}
where the real energy shift and broadening per contact site are analytically given as \cite{dattaElectronicTransportMesoscopic1997,jimenez-valenciaEmergence2ndLaw2026,jimenez-valenciaEntropyFlowQuantum2026}
\begin{align}
\Lambda_p(E) &= \frac{\abs{\tau_w}^2}{2\abs{t_w}^2} \begin{cases}
x+\sqrt{x^2-4\abs{t_w}^2},
&
x<-2\abs{t_w},
\\
x,
&
\abs{x}\leq 2\abs{t_w},
\\
x-\sqrt{x^2-4\abs{t_w}^2},
&
x>2\abs{t_w},
\end{cases} \\
\Gamma_p(E) &=
\begin{cases}
\frac{2\abs{\tau_w}^2}{\abs{t_w}} \sqrt{
1-\left(\frac{x}{2\abs{t_w}}\right)^2},
&
\abs{x}<2\abs{t_w},
\\
0,
&
\abs{x} \geq 2\abs{t_w},
\end{cases} 
\end{align}
respectively, where $x\equiv x(E)=E-\epsilon_w$. Then the total retarded Green's function of Alice's interferometer is
\begin{align}
G^R(E,\phi) &= \left[E-H_{\text{R}}(\phi)+\frac{i}{2}\left(\Gamma^{I}+\Gamma^{II}\right)-\Sigma_p^R(E)\right]^{-1}. \label{full_gf_normal_wire}
\end{align}

We impose the zero-current condition $I_{3}=0$ since the additional normal reservoir is floating. The coherent transmission from terminal $\beta$ to terminal $\alpha$ is
\begin{align}
T_{\alpha\beta}(E,\phi) &= \text{Tr}\left[\Gamma_{\alpha}(E)G^R(E,\phi)\Gamma_{\beta}(E)G^A(E,\phi)\right].
\end{align}
The floating condition in linear response at zero temperature gives \cite{buttikerChemicalPotentialOscillations1989,jimenez-valenciaEmergence2ndLaw2026}
\begin{align}
\mu_3(\phi) &= \frac{T_{31}(E_F,\phi)\mu_1+T_{32}(E_F,\phi)\mu_2}{T_{31}+T_{32}}.
\end{align}
Then the resulting total source to drain transmission is \cite{buttikerCoherentSequentialTunneling1988,jimenez-valenciaPersistentDissipativePeltier2025}
\begin{align}
T_{\text{tot}}(E_F,\phi) &= T_{21}+\frac{T_{23}T_{31}}{T_{31}+T_{32}}. \label{normal_wire_effective_T}
\end{align}
This yields the transmission, contrast, and LDOS at $E=E_F$ of Alice's interferometer coupled to the normal reservoir 
\begin{align}
T(E,\phi) &= T_{\text{tot}}(E,\phi), \\
C(E,\phi) &= \frac{2\abs{T_{\text{tot}}(E,\phi)-T_{\text{tot}}(E,0)}}{T_{\text{tot}}(E,\phi)+T_{\text{tot}}(E,0)}, \label{normal_wire_contrast} \\
\rho(x,E,\phi) &= \braket{x|A(E,\phi)|x}, \label{normal_wire_ldos}
\end{align}
respectively, where $A(E,\phi)$ is the spectral function of the coupled system. 

\begin{figure} 
\centering
\includegraphics[width=\linewidth]{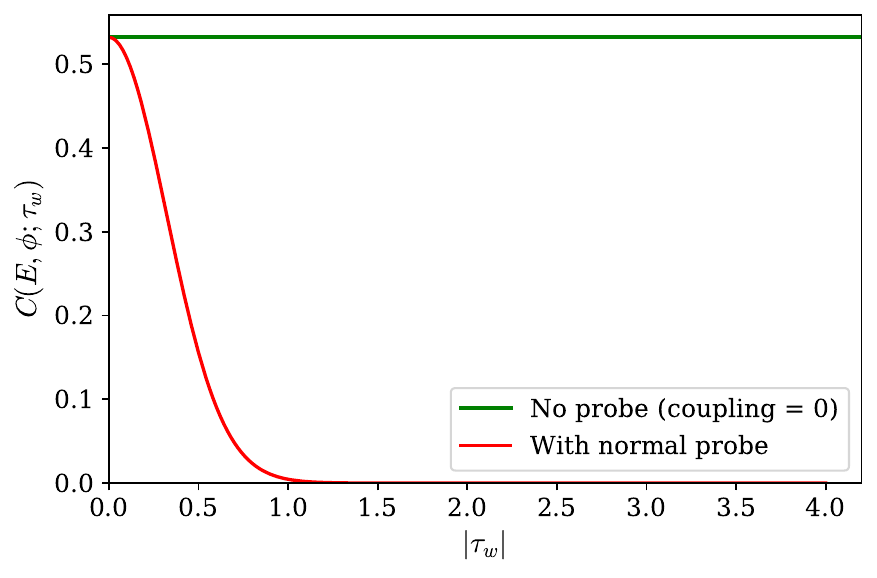}
    \caption{Normal reservoir contrast $C(E,\phi;\tau_{w})$ as a function of $\tau_{w}$ at the band center $E=0$, computed at absolute zero temperature. The contrast decreases monotonically as $\tau_{w}$ increases; there is no reemergence of coherence.}
    \label{fig:contrast_vs_tau_w_semifinite}
\end{figure}

The contrast at the center $E=0$ of the ring's energy band is plotted versus the coupling $\tau_w$ between the ring and the normal reservoir in Fig.\ \ref{fig:contrast_vs_tau_w_semifinite}.
The interferometer parameters used are the same as in the main text, with $E=E_F$, $\epsilon_w=0$, 
$t_w=-1$, and $N_c=10$. 
Fig.~\ref{fig:contrast_vs_tau_w_semifinite} shows that as the coupling $\tau_{w}$ is increased, the interferometer contrast decreases monotonically; there is no reemergence of coherence due to the absence of mode conversion processes.

\clearpage
\onecolumngrid


\makeatletter

\let\section\PRLoriginalsection

\let\@hangfrom@section\PRLoriginalhangfromsection

\let\@sectioncntformat\PRLoriginalsectioncntformat

\let\@runin@to@section\PRLoriginalrunintosection

\makeatother


\clearpage
\onecolumngrid


\makeatletter

\renewcommand{\section}{%
    \@startsection{section}{1}{\z@}%
    {0.8cm \@plus1ex \@minus .2ex}%
    {0.5cm}%
    {\normalfont\small\bfseries\centering}%
}

\let\@sectioncntformat\relax

\def\@hangfrom@section#1#2#3{%
    \@hangfrom{#1#2}\MakeTextUppercase{#3}%
}

\makeatother


\setcounter{section}{0}
\renewcommand{\thesection}{\Roman{section}}

\renewcommand{\theHsection}{supp.\arabic{section}}


\counterwithout{equation}{section}

\renewcommand{\theequation}{S\arabic{equation}}
\setcounter{equation}{0}


\begin{center}
    \textbf{\Large{Supplementary Material for}} \\
    \textbf{\Large{``Decoherence and the Reemergence of Coherence From a Superconducting ``Horizon''''}}
\end{center}

\section{Normal Metal Spacer Tight-Binding Hamiltonian} \label{app:space_ham}

The 2D normal metal spacer's tight-binding Hamiltonian $H_{\text{N}}$ is
\begin{align}
    H_{\text{N}} &= H_{\text{N}}^{\text{on}} + H_{\text{N}}^{(x)} + H_{\text{N}}^{(y)}, \\
    H_{\text{N}}^{\text{on}} &=\varepsilon_{0} \sum^{M_{x}-1,M_{y}-1}_{n,m=0} c_{j(n,m)}^{\dagger} c_{j(n,m)},  \\
    H_{\text{N}}^{(x)} &= t_{x} \sum^{M_{x}-2,M_{y}-1}_{n,m=0} \left(c_{j(n,m)}^{\dagger} c_{j(n+1,m)} + \text{h.c.} \right), \\
    H_{\text{N}}^{(y)} &= t_{y} \sum^{M_{x}-1,M_{y}-2}_{n,m=0} \left(c_{j(n,m)}^{\dagger} c_{j(n,m+1)} + \text{h.c.} \right), 
\end{align}
where $n = 0,1, \ldots, M_{x}-1$, $m = 0,1, \ldots, M_{y}-1$, and $j(n,m) = N+nM_{y}+m$. The spacer-ring 
tight-binding Hamiltonian is
\begin{align}
    H_{\text{N-R}} &= t_{x}^{\prime} \sum^{M_{y}-1}_{m=0} \left(c_{r_{m}}^{\dagger} c_{j(0,m)} + \text{h.c.} \right), 
\end{align}
where $r_{n}$ is the $n$-th ring site of the coupling between the ring and the spacer with $t_{x}^{\prime}$ being the coupling of the spacer and ring. 

\section{Mathematical Connection Between the Low-Energy BdG Theory and Dirac Field Theory}

At subgap energies close to the Fermi surface $\abs{E-E_{F}} < \abs{\Delta}$, the physics of the SN system are governed by the $2\times 2$ Bogoliubov-de Gennes (BdG) equations \cite{bogolyubovTheorySuperfluidity1947,gennesSuperconductivityMetalsAlloys2018}
\begin{align}
    &H_{\text{BdG}} \psi(\mathbf{x},t) = E\psi(\mathbf{x},t), \label{bdg_eqs_0} \\
    &H_{\text{BdG}} =
    \begin{pmatrix}
        H_{0}  & \Delta(\mathbf{x},t) \\
        \bar{\Delta}(\mathbf{x},t)& -H_{0}
    \end{pmatrix}, \label{bdg_ham_0} \\
    &\psi(\mathbf{x},t) = \left(\begin{array}{c} u(\mathbf{x},t) \\ v(\mathbf{x},t) \end{array}\right),\\
    &H_{0} = -\frac{\hbar^{2}}{2m} \nabla_{\perp}^{2} - \mu(\mathbf{x}) + V(\mathbf{x}) = -\frac{\hbar^{2}}{2m} \left(\partial^{2}_{x} + \partial^{2}_{y} \right) - \mu(\mathbf{x}) + V(\mathbf{x}), \label{single part ham 0} \\ &\Delta(\mathbf{x},t) = \Delta_{0}(\mathbf{x},t) e^{i\phi_{S}(\mathbf{x},t)}, \label{order_param_0}
\end{align}
where $\mu(\mathbf{x})$ is the chemical potential. Note that we have promoted the gap $\Delta(\mathbf{x},t)$ to be time and space-dependent. We can decompose the gap $\Delta(\mathbf{x},t)$ as 
\begin{align}
    \Delta(\mathbf{x},t) = \Delta_{1}(\mathbf{x},t) + i\Delta_{2}(\mathbf{x},t), \quad \bar{\Delta}(\mathbf{x},t) = \Delta_{1}(\mathbf{x},t) - i\Delta_{2}(\mathbf{x},t),
\end{align}
where
\begin{align}
    \Delta_{1}(\mathbf{x},t) &= \Re\left(\Delta(\mathbf{x},t) \right)=\Delta_{0}(\mathbf{x},t) \cos(\phi_{S}(\mathbf{x},t)), \\
    \Delta_{2}(\mathbf{x},t) &= \Im\left(\Delta(\mathbf{x},t) \right)=\Delta_{0}(\mathbf{x},t) \sin(\phi_{S}(\mathbf{x},t)).
\end{align}
Then Hamiltonian \eqref{bdg_ham_0} can be written as \cite{volovikSuperconductivityLinesGAP1993}
\begin{align}
    H_{\text{BdG}} &= \vec{h}\cdot\vec{\tau}=H_{0} \tau_{3} + \Delta_{1}(\mathbf{x},t) \tau_{1} - \Delta_{2}(\mathbf{x},t) \tau_{2}, \label{bdg_ham_0_mat_form}
\end{align}
where
\begin{align}
    &\vec{h} =\left(\Delta_{1}(\mathbf{x},t),-\Delta_{2}(\mathbf{x},t),H_{0} \right) , \\
    &\vec{\tau} = \left(\tau_{1}, \tau_{2}, \tau_{3} \right) = \left( 
    \begin{pmatrix}
        0 & 1 \\
        1 & 0
    \end{pmatrix},
    \begin{pmatrix}
        0 & -i \\
        i & 0
    \end{pmatrix},
    \begin{pmatrix}
        1 & 0 \\
        0 & -1
    \end{pmatrix}
    \right) \label{isospin_mats}
\end{align}
are the Hamiltonian vector and $2\times 2$ Nambu matrices, respectively. Note that the Nambu matrices $\vec{\tau}$ are the Pauli matrices except that they act on Nambu space. 

Suppose that we have a constant chemical potential $\mu_{F} = \hbar^{2}k^{2}_{F}/(2m)$. Without loss of generality, we let the potential energy $V$ be constant and set to $V=0$. If $V\neq0$, then $V$ can be absorbed into the ground state energy. Then we have the simplification
\begin{align}
    H_{0} = -\frac{\hbar^{2}}{2m} \nabla_{\perp}^{2} - \mu_{F}.
\end{align}
To study Andreev reflection, we work under the Andreev approximation where $\abs{E-E_F}, \abs{\Delta(\mathbf{x},t)} \ll E_{F}$. Furthermore, we assume that $\Delta(\mathbf{x},t)$ varies slowly over scales of order $k_{F}^{-1}$ \cite{andreevThermalConductivityIntermediate1964}. Now we use the following Ansatz for the two-components of the Nambu spinor $\psi$ 
\begin{align}
    u(\mathbf{x},t) &= f(\mathbf{x},t) e^{i\mathbf{k}_{F}\cdot \mathbf{x}}, \label{u_ansatz} \\
    v(\mathbf{x},t) &= g(\mathbf{x},t) e^{i\mathbf{k}_{F}\cdot \mathbf{x}}, \label{v_ansatz}
\end{align}
where $\mathbf{k}_{F}= k_{F}\hat{\mathbf{k}}_{F}$ is Fermi wave vector and $f$ and $g$ vary on the scale of $k_{F}^{-1}$ as well. If we insert Eqs.~\eqref{u_ansatz} and \eqref{v_ansatz} into Eq.~\eqref{bdg_eqs_0} and ignore terms involving derivatives higher than one, we get the linearized equations \cite{andreevThermalConductivityIntermediate1964}
\begin{align}
    -i\hbar \mathbf{v}_{F} \cdot \nabla_{\perp} f(\mathbf{x},t) + \Delta(\mathbf{x},t) g(\mathbf{x},t) &= i\hbar \frac{\partial}{\partial t} f(\mathbf{x},t), \\
    i\hbar \mathbf{v}_{F} \cdot \nabla_{\perp} g(\mathbf{x},t) + \Bar{\Delta}(\mathbf{x},t) f(\mathbf{x},t) &= i\hbar \frac{\partial}{\partial t} g(\mathbf{x},t), 
\end{align}
where $\mathbf{v}_{F} = (\hbar/m) \mathbf{k}_{F}$ is the Fermi velocity. This yields the linearized BdG Hamiltonian
\begin{align}
    \Tilde{H}_{\text{BdG}} &= -i\hbar \tau_{3}\mathbf{v}_{F} \cdot  \nabla_{\perp}   + \Delta_{1}(\mathbf{x},t) \tau_{1} + \Delta_{2}(\mathbf{x},t) \tau_{2}. \label{linear_bdg_ham}
\end{align}
To express the linearized BdG Hamiltonian \eqref{linear_bdg_ham} in a more mathematically appealing way, we first rewrite the gap terms as
\begin{align}
    \Delta_{1}(\mathbf{x},t) \tau_{1} + \Delta_{2}(\mathbf{x},t) \tau_{2} &=\Delta_{0}(\mathbf{x},t) \left[\cos(\phi_{S}(\mathbf{x},t)) \tau_{1} - \sin(\phi_{S}(\mathbf{x},t)) \tau_{2} \right] = \vec{\Delta}(\mathbf{x},t) \cdot \vec{\tau},
\end{align}
where
\begin{align}
    \vec{\Delta}(\mathbf{x},t) = \Delta_{0}(\mathbf{x},t) \left(\cos(\phi_{S}(\mathbf{x},t)), -\sin(\phi_{S}(\mathbf{x},t)), 0 \right).
\end{align}
This turns Eq.~\eqref{linear_bdg_ham} into
\begin{align}
    \Tilde{H}_{\text{BdG}} &= -i\hbar \tau_{3}\mathbf{v}_{F} \cdot \nabla_{\perp} + \vec{\Delta}(\mathbf{x},t) \cdot \vec{\tau}. \label{linear_bdg_ham_2}
\end{align}
We note that the linearized BdG Hamiltonian \eqref{linear_bdg_ham_2} now closely resembles the Dirac equation except that we have a vector-valued effective ``mass'' term $\vec{\Delta}(\mathbf{x},t)$ that is position and time-dependent. In addition, the effective mass $\vec{\Delta}(\mathbf{x},t)$ rotates in Nambu (pseudospin) space as the order parameter $\phi_{S}(\mathbf{x},t)$ varies.

\section{Disordered Ring} \label{sec:disordered_ring}

We demonstrate that decoherence and the reemergence of coherence persists even in a disordered AB ring coupled to a superconductor. 

The disordered Hamiltonian of our $N$-site tight-binding AB ring is
\begin{align}
    H_{\text{R}}^{(s)}(\phi) &= \sum_{n=0}^{N-1} \left[\left(\varepsilon_{0} + V_{n}^{(s)} \right) c_{n}^{\dagger}c_{n} + t_{0} \left(e^{i\phi/N} c_{n}^{\dagger} c_{n+1} + \text{h.c.} \right) \right], \label{disorder_ab_ring_ham}
\end{align}
where the random potential $V_{n}^{(s)}$ is the independently sampled from a uniform distribution ($V_{n}^{(s)} \sim \mathcal{U}\left(-W\abs{t_{0}},+W\abs{t_{0}} \right)$) where $W \geq 0$ is the dimensionless disorder strength. The superscript $(s)$ labels an independent disorder realization, or seed. More explicitly, the disorder potential can be expressed as
\begin{align}
    V_{n}^{(s)} = 
    \begin{cases}
        \text{independent samples from } \mathcal{U}\left(-W\abs{t_{0}},+W\abs{t_{0}} \right), &\quad n\in\mathcal{D}, \\
        0, &\quad n\notin \mathcal{D},
    \end{cases}
\end{align}
where $\mathcal{D}=\{0,1,\ldots,N-1\}$ are the sites to apply the disorder. Then the disordered Green's function of Alice's interferometer \eqref{disorder_ab_ring_ham} is
\begin{align}
    \Tilde{G}^{R(s)}(E,\phi) &= \left[E - H_{\text{R}}^{(s)}(\phi) +i \left(\Gamma^{I} + \Gamma^{II} \right)/2\right]^{-1},
\end{align}
where it is implicit that $E\equiv E + i\eta$ for the correct pole prescription. This changes the Andreev self-energy to its disordered version
\begin{align}
    \Sigma_{\text{AR}}^{R(s)}(E,\phi) &= \sum_{i,j \in \mathcal{S}} t_{\text{AR}}^{2} \Tilde{G}_{ij}^{R(s)}(E,\phi) \ket{i}\bra{j}.
\end{align}
Then the full retarded disordered Green's function of our model is
\begin{align}
    G^{R(s)}(E,\phi) &= \left[E - H_{\text{R}}^{(s)}(\phi) - \Sigma^{R(s)}(E,\phi)\right]^{-1}, \label{full_gf_disorder}
\end{align}
with the full disordered self-energy
\begin{align}
    \Sigma^{R(s)}(E,\phi) &= -i\frac{\Gamma^{I} + \Gamma^{II}}{2} + \Sigma_{\text{AR}}^{R(s)}(E,\phi).
\end{align}

We see that the inclusion of disorder in Eq.~\eqref{disorder_ab_ring_ham} merely leads to replacing the main text's $G^{R}$ and $\Sigma^{R}$ with $G^{R(s)}$ and $\Sigma^{R(s)}$, respectively. This means that for each \textit{individual} disorder seed $s$, the usual transmission, contrast, and LDOS formulas must change to account for the disorder. In order to achieve this, we replace the main text's Green's function \eqref{superconductor_on_ring_gf} with Eq.~\eqref{full_gf_disorder} to get the disordered transmission, contrast, and LDOS formulas 
\begin{align}
    T^{(s)}(E,\phi) &= \text{Tr}\left[\Gamma^{I} G^{R(s)}(E,\phi) \Gamma^{II} G^{A,(s)}(E,\phi) \right], \\
    C^{(s)}(E,\phi) &= \frac{2 \abs{T^{(s)}(E,\phi) - T^{(s)}(E,0)}}{T^{(s)}(E,\phi) + T^{(s)}(E,0)}, \\
    \rho^{(s)}(x,E,\phi) &= \braket{x| A^{(s)}(E,\phi) |x},
\end{align}
respectively, where
\begin{align}
    A^{(s)}(E,\phi) = \frac{1}{2\pi i}\left[G^{A(s)}(E,\phi) - G^{R(s)}(E,\phi) \right]
\end{align}
is the disordered spectral function. A single disorder seed simulation is not enough to capture the full effects of disorder on our model so we choose to run $N_{s}=100$ seeds. To quantify the results of this ensemble simulation, we use the ensemble averaged transmission, contrast, and LDOS
\begin{align}
    \braket{T(E,\phi)}_{\text{dis}} &= \frac{1}{N_{s}} \sum_{s=1}^{N_{s}} T^{(s)}(E,\phi), \\
    \braket{C(E,\phi)}_{\text{dis}} &= \frac{1}{N_{s}} \sum_{s=1}^{N_{s}} C^{(s)}(E,\phi), \\
    \braket{\rho(x,E,\phi)}_{\text{dis}} &= \frac{1}{N_{s}} \sum_{s=1}^{N_{s}}  \rho^{(s)}(x,E,\phi), 
\end{align}
respectively, of Alice's disordered interferometer. 

For all of our disorder simulations, we use the parameters: $N=100$, $\phi=\pi$, $E_F=0$, $t_{0}=t_{x}=t_{y}=t_{x}^{\prime}=-1$, $\abs{\Delta}=1$, $\gamma =0.2$, rank=20 $\Gamma^{I,II}$ matrices, $\mathcal{S}=\{20,\ldots, 29\}$ (so $M_{y}=\abs{\mathcal{S}}=10$ superconducting sites), and $N_{s}=100$. Note that the ring and source/drain parameters are unchanged from the main text. 

\begin{figure}[H]
  \centering

  \begin{subfigure}[t]{0.85\textwidth}
    \centering
    \includegraphics[width=\linewidth]{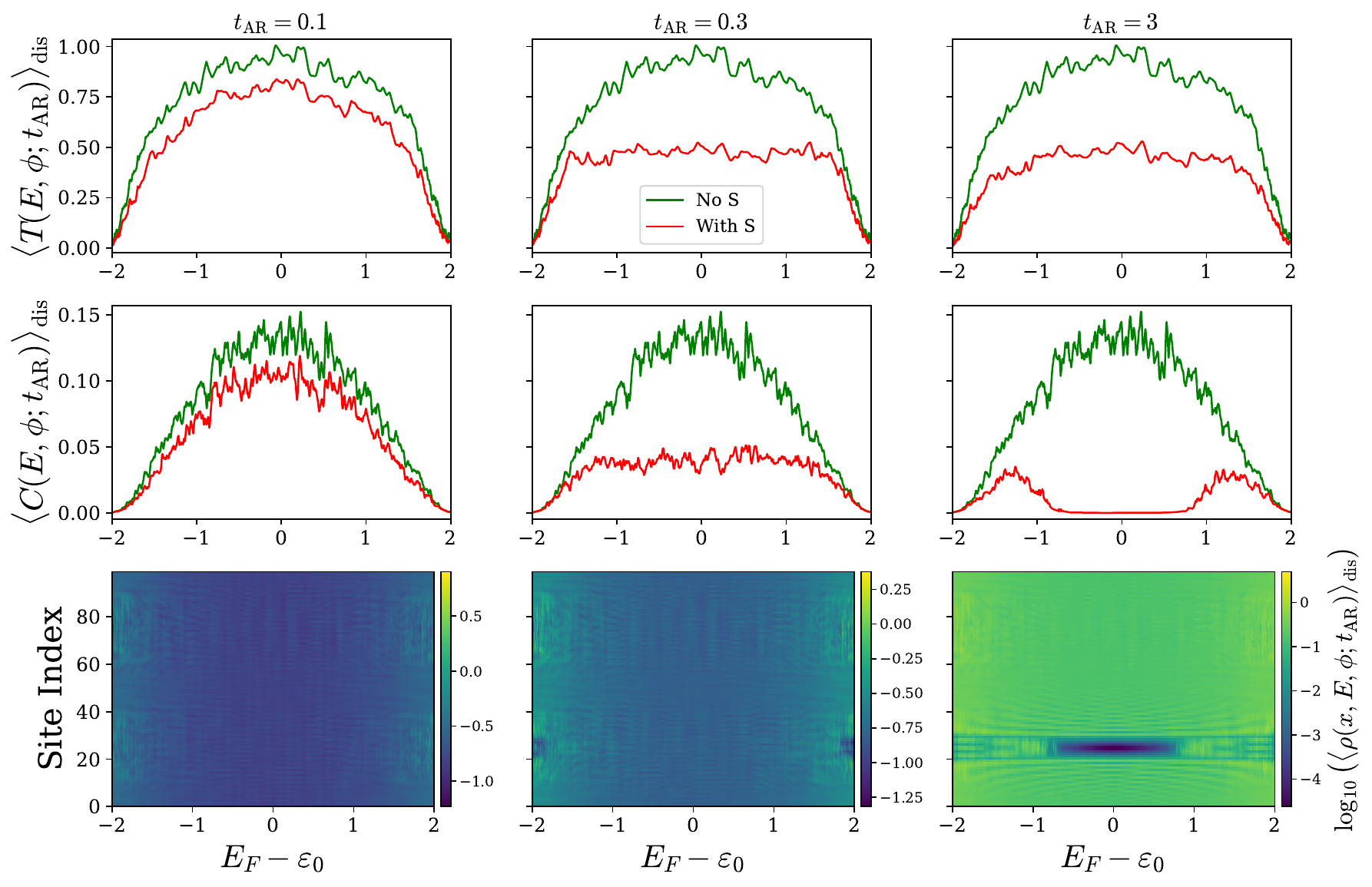}
    \caption{\centering Disorder $W=0.5$}
    \label{fig:ensemble_disorder0.5_3x3}
  \end{subfigure}
  
  \begin{subfigure}[t]{0.85\textwidth}
    \centering
    \includegraphics[width=\linewidth]{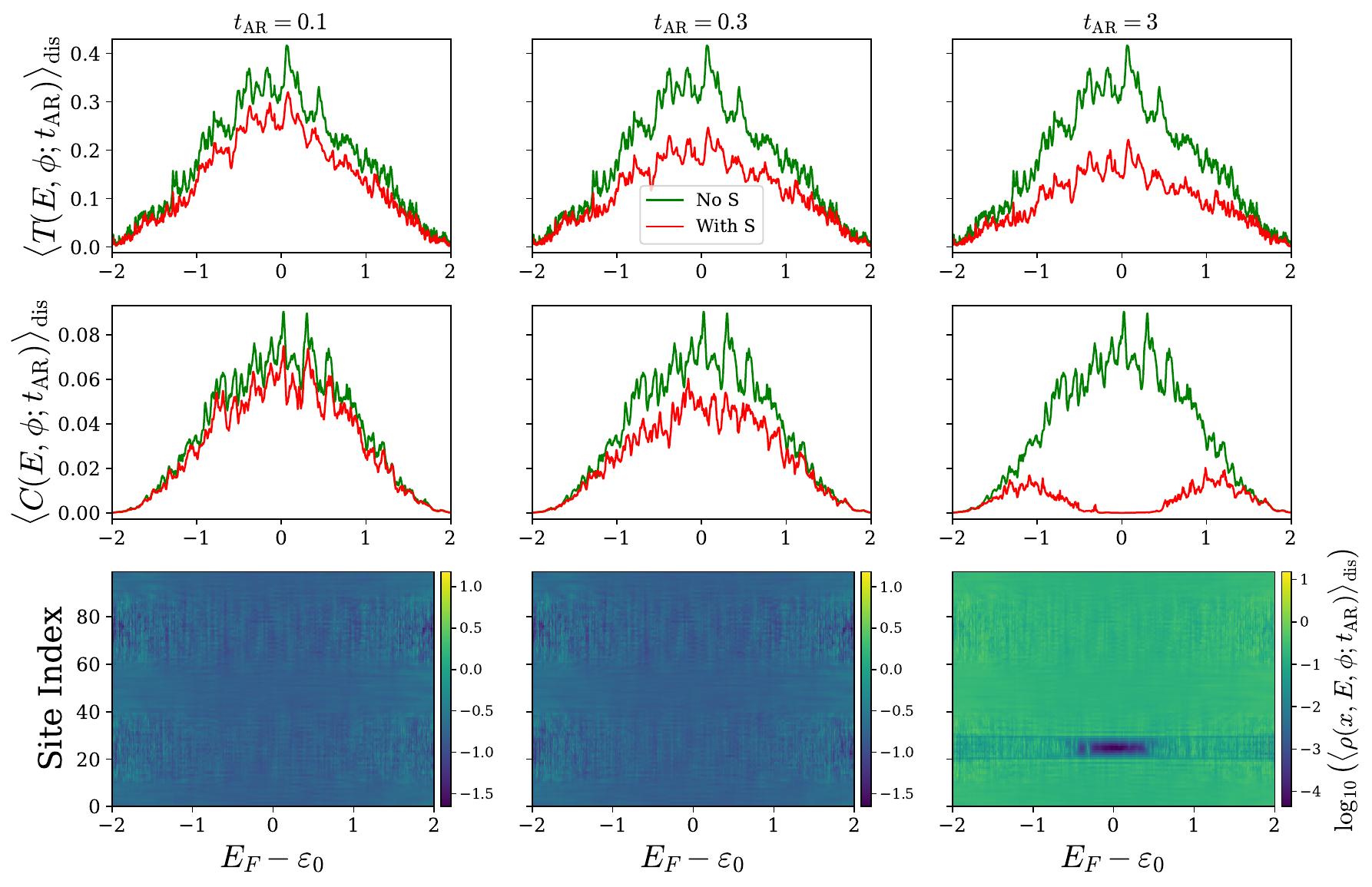}
    \caption{\centering Disorder $W=1$}
    \label{fig:ensemble_disorder1_3x3}
  \end{subfigure}

  \caption{Ensemble disorder averaged transmission $\braket{T(E,\phi; t_{\text{AR}})}_{\text{dis}}$, contrast $\braket{C(E,\phi; t_{\text{AR}})}_{\text{dis}}$, and LDOS $\braket{\rho(x,E,\phi; t_{\text{AR}})}_{\text{dis}}$ computed at absolute zero temperature at $E=E_F$ vs $E_F-\varepsilon_0$ for disorder strength (a) $W=0.5$ and (b) $W=1$ at three values of $t_{\text{AR}}$. See ancillary files for videos illustrating the evolution of $T$, $C$, and $\rho$ as a function of $t_{\text{AR}}$ for disorder strengths $W=0.5$ and $W=1$.}
  \label{fig:TxCxLDOS_disorder}
\end{figure}

We observe in Figs.~\ref{fig:TxCxLDOS_disorder} and the main text's Fig.~(6) that the decoherence \textit{and} the reemergence of coherence is preserved, despite the presence of disorder. In particular, we observe in the main text's Fig.~(6) nearly the same oscillatory and flatline behavior of the contrast as the coupling $t_{\text{AR}}$ increases, corresponding to the reemergence of coherence and decoherence, respectively. The main difference between Figs.~\ref{fig:TxCxLDOS_disorder} and the main text's Fig.~(6) and Figs.~(3) and (4) of the main text are the fluctuations that are present due to the disorder.

Interestingly, we see in the main text's Fig.~(6) that the first reemergence of coherence occurs near $t_{\text{AR}}=0.3$, which is close to the same $t_{\text{AR}}$ value for the non disordered scenario. In realistic scenarios, disorder is present in some form in all materials. Since our findings validate the existence of the reemergence of coherence despite disorder, this leads to the promising possibility of designing gravitational analogue experiments to observe the reemergence of coherence.

\section{Normal Reservoir} \label{sec:normal_reservoir}

It can be argued that other material bodies can reproduce the decoherence and recoherence signatures of our superconductor-normal metal system, much like how other astronomical bodies can reproduce the DSW decoherence \cite{biggsComparingDecoherenceEffects2024}. We show that while some aspects of our solid-state DSW decoherence can be mimicked when replacing the superconductor with a normal metal reservoir, the entire signature is not fully reproduced. In particular, we demonstrate that the reemergence of coherence cannot be replicated by the replacement of the BCS superconductor in our model with a normal reservoir. 

Suppose that each ring site $s_a\in\mathcal{S}$ previously coupled to the superconductor is instead connected to an independent semi-infinite normal tight-binding wire, each of which is connected to normal fermionic reservoir at infinity. The Hamiltonian of wire $a$ and its coupling to the ring are
\begin{align}
H_a &= \sum_{n=0}^{\infty}\left[\epsilon_w d_{a,n}^{\dagger}d_{a,n} + t_w\left(d_{a,n+1}^{\dagger}d_{a,n} + \text{h.c.}\right)\right], \\
H_{\text{R}w,a} &= \tau_w c_{s_a}^{\dagger}d_{a,0} + \tau_w^{*}d_{a,0}^{\dagger}c_{s_a},
\end{align}
respectively, where $\epsilon_w$ and $t_w$ are the onsite energy and nearest-neighbor hopping of the normal wires, and $\tau_w$ is the ring-wire hopping amplitude. Although the $N_c=\abs{\mathcal{S}}$ wires are independent channels, they are coupled to a single floating normal reservoir, denoted by terminal $p=3$.

We analytically integrate out the degrees of freedom of the semi-infinite wires. The retarded surface Green's function of an isolated wire is
\begin{align}
g_{\text{surf}}^R(E) &= \bra{a,0}\left[E-H_a\right]^{-1}\ket{a,0},
\end{align}
where the retarded pole prescription is implicit $E\equiv E+i\eta$. The total retarded self-energy of the floating reservoir is then
\begin{align}
\Sigma_p^R(E) &= \abs{\tau_w}^2 g_{\text{surf}}^R(E)\sum_{a=1}^{N_c}\ket{s_a}\bra{s_a} = \left[\Lambda_p(E)-\frac{i}{2}\Gamma_p(E)\right]\sum_{a=1}^{N_c}\ket{s_a}\bra{s_a},
\end{align}
where $\Lambda_p(E)$ and $\Gamma_p(E)$ are the real energy shift and broadening per contact site, respectively. For a semi-infinite one-dimensional tight-binding wire, these are analytically given as \cite{dattaElectronicTransportMesoscopic1997,jimenez-valenciaEmergence2ndLaw2026,jimenez-valenciaEntropyFlowQuantum2026}
\begin{align}
\Lambda_p(E) &= \frac{\abs{\tau_w}^2}{2\abs{t_w}^2}
\begin{cases}
\left(E-\epsilon_w\right)+\sqrt{\left(E-\epsilon_w\right)^2-4\abs{t_w}^2},
&
\left(E-\epsilon_w\right)<-2\abs{t_w},
\\
E-\epsilon_w,
&
\abs{E-\epsilon_w}\leq 2\abs{t_w},
\\
(E-\epsilon_w)-\sqrt{\left(E-\epsilon_w\right)^2-4\abs{t_w}^2},
&
E-\epsilon_w>2\abs{t_w},
\end{cases} \\
\Gamma_p(E) &=
\begin{cases}
\frac{2\abs{\tau_w}^2}{\abs{t_w}} \sqrt{
1-\left(\frac{E-\epsilon_w}{2\abs{t_w}}\right)^2},
&
\abs{E-\epsilon_w}<2\abs{t_w},
\\
0,
&
\abs{E-\epsilon_w} \geq 2\abs{t_w},
\end{cases} 
\end{align}
respectively. Then the total retarded Green's function of Alice's interferometer is
\begin{align}
G^R(E,\phi) &= \left[E-H_{\text{R}}(\phi)+\frac{i}{2}\left(\Gamma^{I}+\Gamma^{II}\right)-\Sigma_p^R(E)\right]^{-1}. \label{full_gf_normal_wire}
\end{align}

Since the additional normal reservoir is floating, we impose the zero-current condition $I_3=0$. Denoting the transmission from terminal $\beta$ to terminal $\alpha$ by
\begin{align}
T_{\alpha\beta}(E,\phi) &= \text{Tr}\left[\Gamma_{\alpha}(E)G^R(E,\phi)\Gamma_{\beta}(E)G^A(E,\phi)\right],
\end{align}
where $\Gamma_1=\Gamma^{I}$, $\Gamma_2=\Gamma^{II}$, and $\Gamma_3(E)=\Gamma_p(E)\sum_{a=1}^{N_c}\ket{s_a}\bra{s_a}$, the floating condition in linear response at zero temperature gives \cite{buttikerChemicalPotentialOscillations1989,jimenez-valenciaEmergence2ndLaw2026}
\begin{align}
\mu_3(\phi) &= \frac{T_{31}(E_F,\phi)\mu_1+T_{32}(E_F,\phi)\mu_2}{T_{31}+T_{32}}.
\end{align}
Here $\mu_3(\phi)$ is a floating electrochemical potential. The resulting total source to drain transmission is \cite{buttikerCoherentSequentialTunneling1988,jimenez-valenciaPersistentDissipativePeltier2025}
\begin{align}
T_{\text{tot}}(E_F,\phi) &= T_{21}+\frac{T_{23}T_{31}}{T_{31}+T_{32}}. \label{normal_wire_effective_T}
\end{align}
We use this effective transmission to calculate the AB contrast. The total transmission, contrast, and LDOS at $E=E_F$ of Alice's interferometer coupled to the normal reservoir are 
\begin{align}
T(E,\phi) &= T_{\text{tot}}(E,\phi), \\
C(E,\phi) &= \frac{2\abs{T_{\text{tot}}(E,\phi)-T_{\text{tot}}(E,0)}}{T_{\text{tot}}(E,\phi)+T_{\text{tot}}(E,0)}, \label{normal_wire_contrast} \\
\rho(x,E,\phi) &= \braket{x|A(E,\phi)|x}, \label{normal_wire_ldos}
\end{align}
respectively, where
\begin{align}
A(E,\phi) &= \frac{1}{2\pi i}\left[G^A(E,\phi)-G^R(E,\phi)\right]
\end{align}
is the spectral function of the coupled system. 

For all of our normal reservoir simulations, we choose $N=100$, $\phi=\pi$, $E=E_F$, $t_{0}=t_{x}=t_{y}=t_{x}^{\prime}=-1$, $\abs{\Delta}=1$, $\gamma =0.2$, rank=20 $\Gamma^{I,II}$ matrices, $\mathcal{S}=\{20,\ldots, 29\}$ (so $M_{y}=\abs{\mathcal{S}}=10$ superconducting sites), $\epsilon_w=0$, $t_w=-1$, and $N_c=10$. Note that the ring and source/drain parameters are unchanged from the main text. The coupling is varied through $\tau_w$, or equivalently the dimensionless parameter $\lambda_N=\abs{\tau_w/t_0}$. At the wire band center $E=0$, the real shift vanishes and the broadening is $\Gamma_p(0)=2\abs{\tau_w}^2$.

\begin{figure}[H]
\centering
\includegraphics[width=\textwidth]{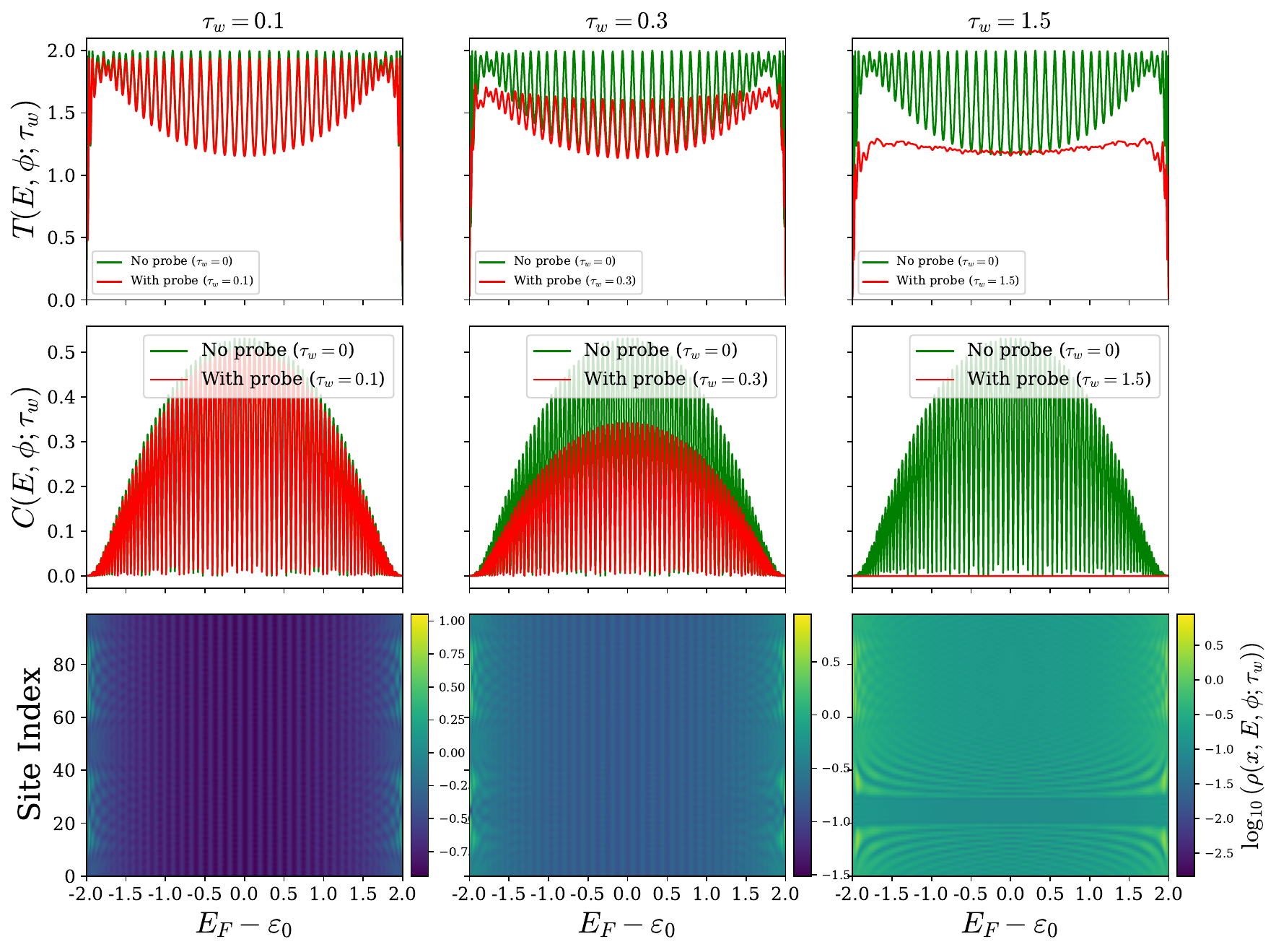}
    \caption{ Transmission $T(E,\phi;\tau_{w})$, contrast $C(E,\phi;\tau_{w})$, and LDOS $\rho(x,E,\phi;\tau_{w})$ with superconductor replaced by a normal metal reservoir for three values of coupling $\tau_{w}$, computed at absolute zero temperature. We only observe monotonic behavior in the decoherence as $\tau_{w}$ increases. See ancillary file for a video illustrating the evolution of $T$, $C$, and $\rho$ as a function of $\tau_{w}$.}
    \label{fig:semiinf_normal_wire_T_C_LDOS_3x3}
\end{figure}

We see in Figs.~\ref{fig:semiinf_normal_wire_T_C_LDOS_3x3} and \ref{fig:contrast_vs_tau_w_semifinite} that as the coupling $\tau_{w}$ is increased, we observe a monotonic decrease in coherence with no reemergence of coherence. This is most prominently seen in Fig.~\ref{fig:contrast_vs_tau_w_semifinite} where we see a monotonic decrease in contrast. This demonstrates that replacing the BCS superconductor in our model with a normal reservoir does not replicate the reemergence of coherence. In fact, we only observe monotonic decoherence. This highlights the fact that searching for suitable black hole analogues requires careful analysis because the analogues \textit{must} have the equivalent features of a real black hole in order to properly construct an analogue system. The fact that the normal reservoir does not possess an intrinsic mode-conversion process like Andreev reflection demonstrates that a normal reservoir is not a proper black hole analogue. 

We note that the decoherence and recoherence behavior of our model with the BCS superconductor is unique due to Andreev reflection. Merely replacing the superconductor with a normal reservoir cannot replicate the full behavior of our original model. This is strongly reminiscent \textit{in spirit} to how the DSW decoherence signature is unique to black holes due to the gravitational interaction unless there exists some extraordinary properties of matter that can mimic this \cite{biggsComparingDecoherenceEffects2024,danielsonLocalDescriptionDecoherence2025}.

\end{document}